\documentclass[a4paper, superscriptaddress, notitlepage, amsmath,amssymb, aps, twocolumn, pre, floatfix, 10pt]{revtex4-2}
\usepackage{graphicx}
\usepackage{bm,bbm}
\usepackage[urlcolor=blue]{hyperref}
\hypersetup{colorlinks=true,allcolors=blue}
\usepackage{amssymb} 

\newcommand{\Elr}[1]{\left\langle #1\right\rangle}
\newcommand{\E}[1]{\langle #1\rangle}

 \newcommand{\rmd}{d}
\newcommand{\rme}{\mathrm{e}}
\newcommand{\rmi}{\mathrm{i}}

\newcommand{\f}[1]{\mathbf{#1}}
\newcommand{\x}{\f x}

\newcommand{\abs}[1]{\left\lvert #1 \right\rvert}

\newcommand{\bsig}{\boldsymbol{\sigma}}
\newcommand{\bSig}{\boldsymbol{\Sigma}}

\newcommand{\bnu}{{\boldsymbol{\nu}}}

\newcommand{\bgam}{\boldsymbol{\gamma}}

\newcommand{\ps}{p_{\rm s}}

\newcommand{\var}{{\rm var}}

\newcommand{\cov}{{\rm cov}}

\newcommand{\DS}{\Delta S_{\rm tot}} 
\newcommand{\DTS}{\Delta (TS_{\rm tot})} 

\usepackage{xcolor}
\definecolor{C0}{rgb}{0.12156862745098039, 0.4666666666666667, 0.7058823529411765}
\definecolor{C1}{rgb}{1.0, 0.4980392156862745, 0.054901960784313725}
\definecolor{C2}{rgb}{0.17254901960784313, 0.6274509803921569, 0.17254901960784313}
\definecolor{C2_darker}{rgb}{0.0.14235, 0.517647, 0.0.14235}
\definecolor{C3}{rgb}{0.8392156862745098, 0.15294117647058825, 0.1568627450980392}
\definecolor{C4}{rgb}{0.5803921568627451, 0.403921568627451, 0.7411764705882353}
\definecolor{C5}{rgb}{0.5490196078431373, 0.33725490196078434, 0.29411764705882354}
\definecolor{inferno0}{rgb}{0.087411, 0.044556, 0.224813}
\definecolor{inferno1}{rgb}{0.379001, 0.076253, 0.432719}
\definecolor{inferno2}{rgb}{0.658463, 0.178962, 0.372748}
\definecolor{inferno3}{rgb}{0.894305, 0.353399, 0.193584}
\definecolor{inferno4}{rgb}{0.987622, 0.64532 , 0.039886}

\colorlet{mylinkcolor}{blue!66!black!80}
\hypersetup{colorlinks = true, linkcolor = {mylinkcolor}, linkcolor = {mylinkcolor}, citecolor = {mylinkcolor}, urlcolor = {mylinkcolor}}
\definecolor{grey}{rgb}{0.6,0.6,.6}
\definecolor{darkgrey}{rgb}{0.4,0.4,.4}
\definecolor{darkgreen}{rgb}{0,0.4,0}
\definecolor{lightgreen}{rgb}{0,0.7,0}
\definecolor{darkred}{rgb}{0.5,0,0}
\newcommand{\red}[1]{{\color{red}#1}}

\newcommand{\blue}[1]{{\color{blue}#1}}

\renewcommand{\blue}[1]{{#1}}
\renewcommand{\red}[1]{{#1}}
\allowdisplaybreaks

\begin{document}
\title{Thermodynamic bounds on generalized
  transport:~From single-molecule to bulk observables}
%
\author{Cai Dieball}
\author{Alja\v{z} Godec}
\email{agodec@mpinat.mpg.de}
\affiliation{Mathematical bioPhysics Group, Max Planck Institute for Multidisciplinary Sciences, 37077 G\"ottingen, Germany}

\begin{abstract}
We prove that the transport of any \red{differentiable }scalar observable in $d$-dimensional
 non-equilibrium systems is bounded from above by the total entropy
 production scaled by the amount the
 observation ``stretches'' microscopic coordinates.~The result---a time-integrated
generalized speed limit---reflects the thermodynamic cost of transport
of observables, 
and places underdamped
 and overdamped stochastic dynamics on equal
 footing 
 with deterministic motion.~Our work allows for
stochastic thermodynamics to make contact with bulk experiments,
 and fills an important gap in thermodynamic inference,
 since
 microscopic dynamics is, at least for short times, underdamped.~Requiring only averages but not sample-to-sample
 fluctuations, the proven transport bound is practical and
 applicable not only to single-molecule but also bulk
 experiments where only averages are observed, which we demonstrate by
 examples.~Our results may facilitate
 thermodynamic inference on molecular machines without an
 obvious
 directionality from 
bulk observations of transients probed, e.g.\ in time-resolved X-ray
scattering.
\end{abstract}
\maketitle

A complete thermodynamic
characterization and understanding of systems driven far from
equilibrium remains elusive.~Central to non-equilibrium thermodynamics
is the total entropy production $\Delta S_{\rm tot}$, which reflects
the displacement from equilibrium \cite{Seifert2012RPP}.~$\Delta S_{\rm
  tot}$  embodies the entropy change in both, the system and the
environment coupled to it, and is a measure of the
violation of time-reversal symmetry
\cite{Seifert2005PRL,Seifert2012RPP}.~\blue{Most importantly, the
 \emph{thermodynamic cost} of non-equilibrium processes (stationary,
 transient, or even explicitly time-dependent) at temperature $T$ is
 in fact $T\DS$, which can thus be seen as the ``counterpart'' of free
 energy in equilibrium. The total entropy production allows, for
 example, to quantify the efficiency of molecular motors
 \cite{Seifert2018PA,Leighton2022PRL,Leighton2023PRL} and gain
 insight into the energetic budget of human red blood cells
 \cite{DiTerlizzi2024S}. While most works focus on non-equilibrium
 steady-state (NESS) dynamics, transient processes that approach
 equilibrium states such as, e.g., protein folding \cite{Sato2023,Roessle2000JAC,Akiyama2002PNASU,Stamatoff1979BJ,Yamada2012B,Cho2021COSB,Arnlund2014,Cho2010,Cammarata2008} or thermal
 relaxation \cite{Ibanez2024NP}, are also characterized by $T\DS$ which then corresponds
 to an excess free energy \cite{Ibanez2024NP,Lebowitz1957AP,Vaikuntanathan2009EEL,Lapolla2020PRL}. Even more
 involved are transients towards NESS, relevant in e.g.,
the packaging of viral DNA \cite{Berndsen2014PNASU}, red blood cell
flickering \cite{Turlier2016NP}, enzymatically facilitated topological
relaxation of DNA \cite{Stuchinskaya2009JMB}, or nanoparticle model systems
for interrogating the  fundamental laws of stochastic
thermodynamics \cite{Gieseler2018E}.}

Despite its importance, the
inference of $\Delta S_{\rm tot}$ from experimental observations is
far from simple, as it requires access to \emph{all} dissipative
degrees of freedom is the system, which is typically precluded by the
fact that one only has access to some observable.~Notably, neither
the microscopic dynamics nor the projection underlying the
observable are typically known, and often one can only observe
transients. 
 
To overcome these intrinsic limitations of experiments, diverse
bounds (i.e., inequalities) on the entropy production have been
derived, in particular thermodynamic uncertainty relations (TURs)
\cite{Barato2015PRL,Gingrich2016PRL,Gingrich2017JPAMT,VanVuPRE2019,Horowitz2019NP,Manikandan2020PRL,Otsubo2020PRE,Liu2020PRL,Koyuk2020PRL,Dieball2022PRL,Dieball2022PRR,Dechant2021PRR,Dechant2021PRX,Dieball2023PRL}
and speed limits
\cite{Brandner2013PRL,Shiraishi2018PRL,Shiraishi2019PRL,Massi,Ito_det,CSL_4,CSL_6,Brandner2021PRX,Dechant2018PRE,Ito2020Stochastic,VanVu2023PRX}.
Such bounds provide conceptual insight about manifestations of
irreversible behavior, and from a practical perspective they allow
to infer a bound on $\Delta S_{\rm tot}$ from measured trajectories,
more precisely from the sample-to-sample fluctuations or speed of
observables.\\
\indent These results  remain incomplete from several perspectives. First, their
validity typically hinges on the assumption that the
microscopic dynamics is overdamped or even a Markov-jump
process. This is unsatisfactory because microscopic dynamics is, at
least on short time scales, underdamped and the TUR does \emph{not} hold
for underdamped dynamics \cite{Patrick_break}~(see, however, progress in
\cite{Fischer2020PRE,Kwon2022NJP,Fu2022PRE,Lee2023PRR} \blue{and recently \cite{arxiv_Crutchfield_TUR}}). Similarly,
thermodynamic speed limits have so far seemingly not been derived for
underdamped dynamics.
\blue{While inertial effects may not be important in colloidal systems
  \cite{Dhont1996}, they 
  are indeed relevant for, e.g., protein dynamics
  \cite{McCammon1984RPP,Smith1991QRB,Kneller2000,Brunig2022JPCB}, and
  are known to invalidate overdamped theories even on long time scales
  \cite{Patrick_break}. Even the basic formulation of path-wise
  thermodynamics of inertial 
  systems is fundamentally different, and in the analysis of densities
  or integrals over stochastic trajectories one cannot eliminate
  short-time inertial effects.} 
Second, dissipative processes are often, 
especially in molecular machines without an obvious directionality
(e.g.\ molecular chaperones \cite{Chaperones}), mediated by 
intricate collective (and often fast) open-close motions visible in
transients that are
difficult to resolve even with advanced single-molecule techniques
\cite{Thorsten}. Experiments providing more 
detailed structural information, such as time-resolved X-ray scattering
techniques
\cite{Sato2023,Roessle2000JAC,Akiyama2002PNASU,Stamatoff1979BJ,Cho2021COSB,Arnlund2014,Cho2010,Cammarata2008},
are available, but probe bulk behavior for which
the existing bounds do not apply.
There is thus a pressing need to close the gaps and to cover underdamped
dynamics and tap into bulk observations.\\
\indent Here, we present thermodynamic bounds on the generalized transport of
observables in systems evolving according to (generally time-inhomogeneous) overdamped, underdamped,
or even deterministic dynamics, all treated on an equal
footing, \red{generalizing existing results on transport \cite{Leighton2022PRL,Dechant2018PRE}.} Technically, the results may be 
classified as a time-integrated version of generalized speed limits
and bring several conceptual and practical advantages. As a
demonstration, we use the
bounds for thermodynamic inference based on both, single-molecule and bulk
(e.g.\ scattering) observables.\\
\indent 
\emph{Rationale.---}Consider the simplest case
of a Newtonian particle with position $x_\tau$ and velocity $v_\tau$
at time $\tau$ dragged 
through a viscous
medium against the (Stokes) friction force $F_{\gamma}=-\gamma
v_\tau$ with friction constant $\gamma$ causing a transfer of energy into the medium. The dissipated heat
between times $0$ and $t$
is $\Delta Q_{\gamma}=\gamma\int_0^t
v_\tau\rmd x_\tau=\gamma\int_0^tv_\tau^2\rmd\tau$ and gives rise to
entropy production in the medium 
\cite{Seifert2012RPP}. Since deterministic dynamics does not
produce entropy otherwise, we have $T\DS=\Delta Q_{\gamma}$ in
$\tau\in [0,t]$. This imposes a thermodynamic bound on transport via
the Cauchy-Schwarz inequality 
$(x_t-x_0)^2\!=\!(\int_0^tv_\tau1\rmd\tau)^{2}\!\le\!
\int_0^tv_\tau^2\rmd\tau\int_0^t1^2\rmd\tau'$,
yielding $T\DS\ge \gamma (x_t-x_0)^2/t$  
with equality for constant velocity. Therefore, for given $t$
and $\gamma$ a minimum energy input $\Delta Q_\gamma$ is required to achieve a
displacement $x_t-x_0$. The intuition that transport requires dissipation 
extends to
general dynamics 
and scalar observables
as follows.\\
\indent 
\emph{Main result.---}The transport of any \red{differentiable }scalar observable $z_\tau\equiv z(\x_\tau,\tau)$ \blue{(see \cite{Note1} for $z_\tau\equiv z(\x_\tau,\f v_\tau,\tau)$)} on a time interval $[0,t]$ in $d$-dimensional
generally underdamped and time-inhomogeneous dynamics $(\x_\tau,\f
v_\tau)$ is bounded from above by $T\DS$ as
\begin{align}
&T\DS \ge \frac{\Elr{z_t-z_0-\int_0^t\partial_\tau z_\tau\rmd\tau}^2}{t\mathcal{D}^z(t)}\nonumber\\
&\mathcal{D}^z(t)\equiv\frac1t\int_0^t\E{[\nabla_\x z_\tau]^T\bgam^{-1}(\tau)\nabla_\x z_\tau}\rmd\tau,
\label{transport bound} 
\end{align}
where $\bgam$ is a positive definite, possibly time-dependent,
symmetric friction matrix, $\langle\cdot\rangle$ denotes
an ensemble average over non-stationary trajectories,
and $\mathcal{D}^z(t)$ is a
fluctuation-scale function of the observable that determines how
much the observation $z$ \emph{stretches} microscopic coordinates $\x$.  While $\DS$ for
stochastic dynamics differs from $\Delta Q_{\gamma}/T$,
the bound~\eqref{transport bound} remains
valid in the whole spectrum from Newtonian to
underdamped and overdamped stochastic dynamics.~The inequality saturates for
$\nabla_\x z_\tau=c\bgam\bnu(\x,\f v,\tau)$ for any constant $c$ and
$\bnu$ defined in Eq.~\eqref{local mean velocity}. \blue{Note that the application of Eq.~\eqref{transport bound} does \emph{not} require knowledge of the parameters of the underlying motion. One only needs to infer the average in the numerator and $\mathcal D^z$. To infer the latter, different strategies are presented towards the end of the Letter.}


\blue{Setting $z_\tau=x_\tau$ for $d=1$, Eq.~\eqref{transport bound} includes
the above deterministic case. In the overdamped limit, Eq.~\eqref{transport bound} complements the Benamou-Brenier formula \red{that bounds transport in terms of a Wasserstein distance}  \cite{Benamou2000NM,VanVu2023PRX}, and special cases of the bound correspond to existing overdamped speed limits \cite{Dechant2018PRE,Leighton2022PRL,Leighton2023PRL} \red{(in particular, Ref.~\cite{Dechant2018PRE} already contains an overdamped version of $\mathcal D^z$, see \cite{Note1} for a detailed connection to the existing literature)}.}

The bound~\eqref{transport
  bound} characterizes the thermodynamic cost of transport and may be
employed in 
thermodynamic inference. By only requiring the mean but not
sample-to-sample fluctuations, the bound~\eqref{transport bound} is
simpler than the TUR and allows to infer $\DS$ from transients of
bulk observables, probed, e.g., in time-resolved scattering
experiments \cite{Sato2023,Roessle2000JAC,Akiyama2002PNASU,Stamatoff1979BJ,Cho2021COSB,Arnlund2014,Cho2010,Cammarata2008,Josts2020S}. A disadvantage of this simplicity is that it is not
useful for stationary states\red{, unless they are translation invariant or periodic}.
The observable $z_\tau$ can represent a measured projection, whose
functional form 
may be known (e.g., in $X$-ray scattering) or unknown (e.g., a
reaction coordinate of a complex process).~For optimization of thermodynamic
inference $z_\tau$ may be chosen
\emph{a posteriori} and $\tau$-dependent. 

\emph{Outline.---}First we describe the
setup and discuss
different notions of $\DS$ from deterministic via underdamped to
overdamped dynamics. Next we present examples in the context of single-molecule versus bulk X-ray
scattering experiments, as well as higher-order transport
in stochastic heat engines. We then explain how to interpret and infer
the fluctuation-scale function
$\mathcal{D}^z(t)$. We conclude with a perspective.

\emph{Setup.---}Let $\bgam,\f m$ be ${d\times d}$ positive definite, symmetric
friction and mass matrices with square root 
$\sqrt{\bgam}\sqrt{\bgam}^T\!\!=\!\bgam$.~The full dynamics ${\x_\tau,\!\f
v_\tau\!\in\!\mathbbm{R}^d\!}$ evolve according to \cite{Risken1989}
\begin{align}
\rmd\x_\tau &= \f v_\tau\rmd\tau\label{underdamped SDE} \\
\rmd\f v_\tau &= \f m^{-1}\left[\f F(\x_\tau,\tau)\rmd\tau-\bgam\f v_\tau\rmd\tau+\sqrt{2k_{\rm B}T\bgam}\rmd\f W_\tau\right],\nonumber
\end{align}
which in the overdamped limit reduce to 
\begin{align}
\rmd\x_\tau^{\rm od} &= \bgam^{-1}\f F(\x_\tau^{\rm od},\tau)\rmd\tau+\sqrt{2k_{\rm B}T\bgam^{-1}}\rmd\f W_\tau,\label{overdamped SDE} 
\end{align}
where $\f F(\x,\tau)$ is a force field and $\f W_\tau$ the
$d$-dimensional Wiener process. We allow 
$\bgam(\tau)$ (and later also $T(\tau)$) to 
depend on time but suppress this dependence to simplify notation.
We define the \emph{local mean velocity} $\bnu$ of the probability
density $\rho$ in phase space as $\bnu^{\rm Newton}=\f v$ and 
\begin{align}
&\bnu(\x,\f v,\tau)
\equiv\f v+\f m^{-1}k_{\rm B}T\frac{\nabla_{\f v}\rho(\x,\f v,\tau)}{\rho(\x,\f v,\tau)}\nonumber\\
&\bnu^{\rm od}(\x,\tau)
\equiv\bgam^{-1}\left[\f F(\x,\tau)-k_{\rm B}T\frac{\nabla_{\f x}\rho^{\rm od}(\x,\tau)}{\rho^{\rm od}(\x,\tau)}\right].\label{local mean velocity}
\end{align}
The definition for overdamped dynamics is standard
\cite{Seifert2012RPP}, whereas in the underdamped setting $\bnu$ is only the  ``irreversible'' part of the probability current divided by density \cite{Kwon2022NJP}.
Note that $\f v_\tau=\rmd\x_\tau/\rmd\tau$ does not exist for
overdamped dynamics as 
${\E{\abs{\rmd\x_\tau/\rmd\tau}}\to \infty}$ for $\rmd\tau\to 0$ but
$\E{\rmd\x_\tau}/\rmd\tau=\bgam^{-1}\f F(\x,\tau)$ is
well-behaved. The overdamped limit is loosely speaking
obtained for 
$\bgam^{-1}\f m\to\f 0$,
whereby details of this limit depend on $\f F$ \cite{Wilemski1976}.

There are two differences between Newtonian and stochastic dynamics:
(i) the energy exchange between system and bath counteracts friction, and (ii) changes
in $\rho$ give rise to a change in Gibbs entropy of a stochastic system, thus
contributing to the total entropy production as $\DS=\Delta S_{\rm
  sys}+\Delta S_{\rm bath}$. 
In all cases considered we can write the total entropy production as \cite{Seifert2005PRL,Seifert2012RPP,Kwon2022NJP}
\begin{align}
T\DS\!=\!\int_0^t\!\rmd\tau\!\Elr{\bnu^T(\x_\tau,\!\f
  v_\tau,\!\tau)\bgam\bnu(\x_\tau,\!\f v_\tau,\!\tau)}\ge 0\,.\label{S_tot formula}
\end{align}
Within this setup, an \emph{educated guess} and stochastic calculus
alongside the Cauchy-Schwarz inequality delivers the announced
bound~\eqref{transport bound} (\blue{see proof in \cite{Note1}}).   

\begin{figure}
\includegraphics[width=.475\textwidth]{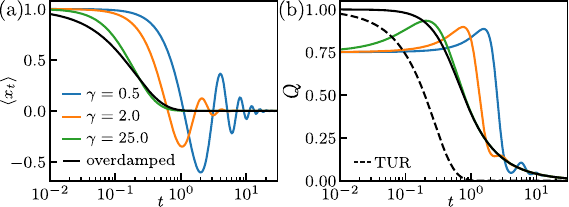}
\caption{Particle $x_t$ in a harmonic trap displaced from $x=1$ to $x=0$ at time
  $0$.~(a)~The particle's mean position $\E{x_0}=1$ moves towards the
  new center of the trap, whereby oscillations occur for small
  damping. The probability density around $\E{x_\tau}$ in this example
  is a Gaussian of constant width (see \cite{Note1} for details and parameters). (b) Quality factor $Q$ of the transport bound for the simplest
 observable $z_\tau=x_\tau$, and quality factor of the TUR for the
 current $J_t=x_t-x_0$ in the overdamped case. Full saturation $Q=1$ at all times can be achieved for
 overdamped and underdamped dynamics as described below.}
\label{fig:trap}
\end{figure}
\emph{Example 1:~Colloid in displaced trap.---}
Consider a bead trapped in a harmonic potential
displaced from position $1$ to $0$ at time $\tau=0$. Knowing
$\gamma$ and observing only the mean particle transport $\E{x_t-x_0}$ we infer the
entropy production from Eq.~\eqref{transport bound} to be $T\DS\ge
T\DS^{\rm bound}\equiv\gamma\E{x_t-x_0}^2/t$.
We inspect the quality of the bound, $Q\equiv \DS^{\rm bound}/\DS\in[0,1]$, as a function of time (see Fig.~\ref{fig:trap}). For
underdamped dynamics $Q$ tends to
$3/4$ at short times due to inertia (see
 \footnote{See Supplemental Material at [...] including references \cite{Dechant2022JPA,Gardiner1985,Pigolotti2017PRL,Dechant2018JSMTE}.} for derivation). Using
$z_\tau=x\nu(\tau)$ (for this example, 
$\nu$ turns out to be independent of $x,v$) we achieve saturation
for all times, which is easily understood from our proof \blue{(see
  \cite{Note1})}. Saturation for this example was also
achieved via the transient correlation TUR~\cite{Dieball2023PRL}.~However, the
present approach is expected to be
numerically more stable and requires less statistics \blue{(see also Fig.~S1 in \cite{Note1})}, since a
determination of
fluctuations and derivatives of observables is not required. Moreover, the
simplest version $z_\tau=x_\tau$
outperforms the transient TUR for the simplest current
$J_t=x_t-x_0$ (dashed line in Fig.~\ref{fig:trap}b). This may be
interpreted as the magnitude of $x_t-x_0$
entering~\eqref{transport bound} being more relevant than its precision entering the TUR for
this example. In contrast to the TUR \cite{Patrick_break}, we also have
the advantage that Eq.~\eqref{transport bound} holds for underdamped
dynamics.

A disadvantage of Eq.~\eqref{transport bound} is that it is not useful for steady-state
dynamics, since there $\E{z_t-z_0}=0$. An exception are
\blue{spatially periodic systems} treated as \blue{NESS} 
\cite{Seifert2018PA,Leighton2022PRL}. A particular
example are \blue{overdamped} Brownian clocks
\cite{Barato2016PRX} where for given $\DS$ the TUR limits
\emph{precision} whereas Eq.~\eqref{transport bound} limits
the magnitude of transport, i.e., the size of the clock. 
\blue{For a quantitative periodic example including comparisons to TURs see Fig.~S1 in \cite{Note1}.}

\emph{Example~2:~Scattering experiments.---}Since~Eq.~\eqref{transport bound} only requires averages, it is applicable beyond single-molecule probes to bulk
experiments, i.e., experiments on samples of many molecules
probing mean properties, e.g., scattering
techniques.~The recent surge in the development
of time-resolved X-ray scattering on proteins
\cite{Sato2023,Roessle2000JAC,Akiyama2002PNASU,Stamatoff1979BJ,Yamada2012B,Cho2021COSB,Arnlund2014,Cho2010,Cammarata2008}
renders our bound
particularly useful. 
Here, transients may be excited by a pressure \cite{Dave2015CMLS,Woenckhaus2001BJ} or
temperature \cite{Kubelka2009PPS} quench, or one directly monitors
slow kinetics \cite{Vestergaard2007PB}. One typically observes the structure
factor $S(\f q)\equiv\frac1N\sum_{j,k=1}^N\E{\rme^{-\rmi\f
    q\cdot(\f r_t^{\,j}-\f r_t^{\,k})}}$ \cite{Dhont1996,Dieball2022NJP}, where
the sum runs over all scatterers (atoms, particles, etc.).
This also applies to interacting colloid suspensions,
where $S(\f q)$ is
the Fourier transform of the pair
  correlation function \cite{Dhont1996}.
An even
simpler observable is
the radius of gyration, $R_g^2\equiv\frac1N\sum_{j=1}^N\E{(\f
  r_j-\bar{\f r})^2}$ \cite{Dhont1996,Svergun2003RPP,Yamada2012B}, where $\bar{\f r}\equiv\frac1N\sum_{j=1}^N\f
r_j$ is
the center of mass. $R_g^2$ reflects the (statistical) size of molecules and is
easily inferred from small $q$ via Guinier's law,
${S(|\f q|)\!\overset{|\f q|\to 0}{=}\!S(0)\rme^{-|\f q|^2R_g^2/3}}$ \cite{Dhont1996,Svergun2003RPP}.

We consider the structure factor
averaged over 
spatial dimensions $S(q)$ (see  \cite{Note1} for the vector
version). For simplicity assume that $\gamma$ is a known scalar.
We observe how $S(q)$ changes over time. From Eq.~\eqref{transport
  bound} we can derive the bounds (see \cite{Note1})
\begin{align}
  T\DS&\!\ge\!\frac{3\gamma[S_t(q)-S_0(q)]^2}{q^2\int_0^t[N-S_\tau(2q)]\rmd\tau}
\!\ge\!\frac{3\gamma[S_t(q)-S_0(q)]^2}{q^2t\max_\tau[N-S_\tau(2q)]} \nonumber\\
T\DS&\!\ge\!\frac{N[R_g^2(t)-R_g^2(0)]^2}{4\int_0^t\rmd\tau R_g^2(\tau)/\gamma}\!
\ge\!\frac{N[R_g^2(t)-R_g^2(0)]^2}{4t\max_\tau[R_g^2(\tau)]/\gamma}.\label{bounds Rg Sq}
\end{align}
\begin{figure}
\includegraphics[width=.475\textwidth]{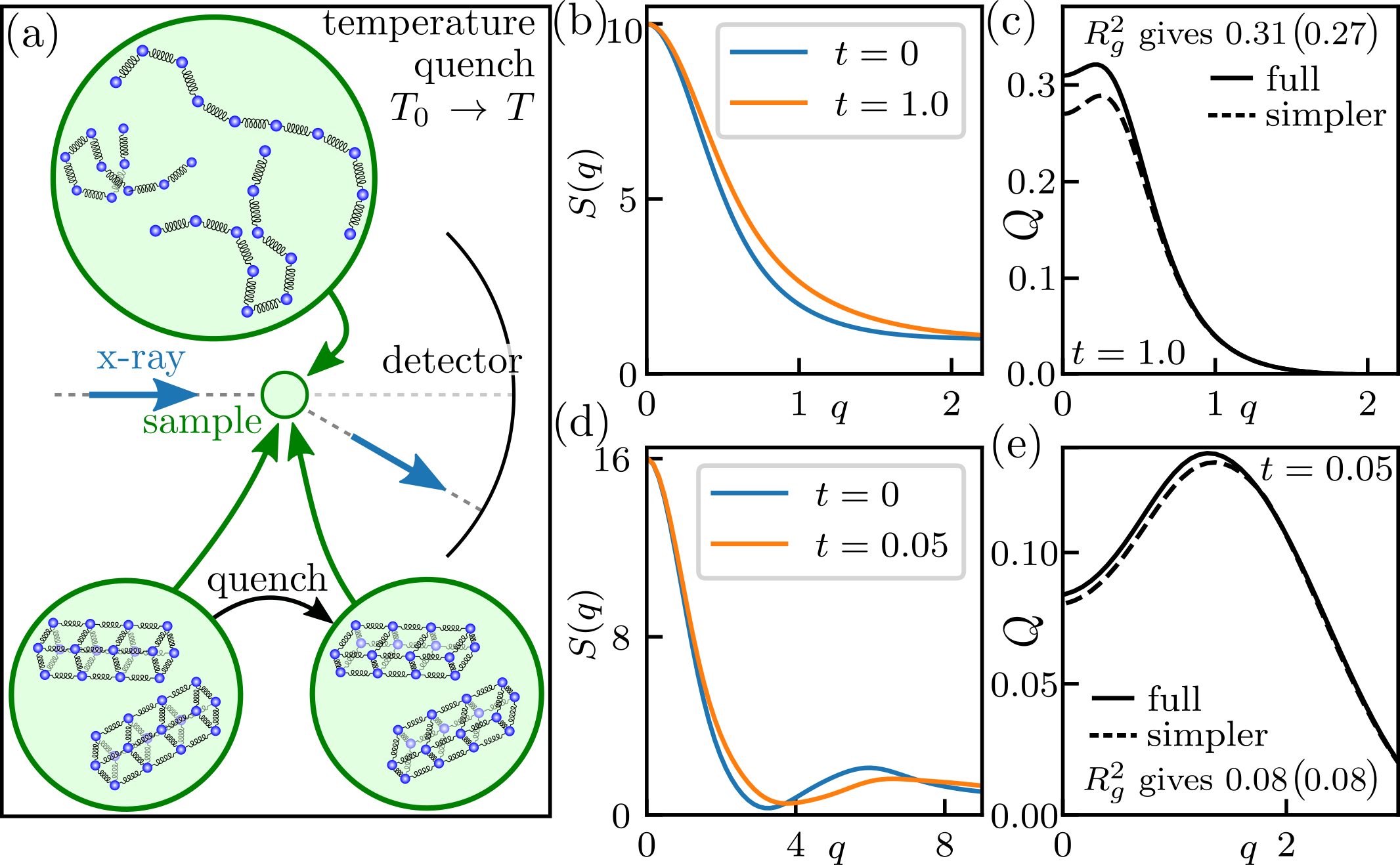}
\caption{(a) Sketch of a scattering setup with (b,c) Rouse polymers with
  $N=10$ beads subject to a temperature quench  from $T_0=2T$ to $T$ at
  $t=0$, and (d,e) harmonically confined ``nano-crystal'' with $N=16$ with Hookean
  neighbor-interactions subject to a quench in rest positions (see \cite{Note1} for model details and parameters). (b,d) Structure factors and (c,e) corresponding quality factors [see Eq.~\eqref{bounds Rg Sq}; ``simpler'' bound contains $\max_\tau$ instead of integral].}
\label{fig:scattering}
\end{figure}
The second bound in each line is a simplification when the
maximum 
over time is known, and it is not necessary to measure
for all $\tau$. The quality of these
bounds for entropy production of internal degrees of freedom during 
relaxation is shown in Fig.~\ref{fig:scattering} for a solution of
(b,c) Rouse polymers upon a temperature quench 
(\red{see Fig.~\ref{fig:scattering}b,c}; probing, e.g., the thermal relaxation asymmetry 
\cite{Lapolla2020PRL,Dieball2023PRR,Ibanez2024NP}) and confined nano-crystals upon a
structural transformation \red{(see Fig.~\ref{fig:scattering}d,e)}, respectively. Dissipative processes
occur on distinct length scales, leading to changes of $S_t(q)$ at
distinct $q$ (Fig.~\ref{fig:scattering}b,d). \red{In both cases, the} sharpness of the bounds~\eqref{bounds Rg Sq} depends on $q$,
giving insight into the participation of distinct modes in
dissipation. The $q$-dependence can in turn be used for optimization
of inference. For small $q$ modes we recover the bound in terms of $R^2_g(\tau)$.
We recover $\sim 30\;\%$ of
$\DS$ in the quenched polymer solution and $\sim 15\;\%$ for the
transforming nano-crystal. This is in
fact a lot \blue{considering that} we only measure a 1d projection
of $3(N-1)$ 
internal
degrees of freedom. \blue{Note that here the large contribution from low
  $q$-modes is intuitive, since the temperature change mostly affects the overall statistical size of the polymer.}\\
\indent
The bound~\eqref{transport bound} will be useful for many bulk experiments
beyond scattering. Consider, for example, a bulk measurement of the mean
FRET efficiency for a pair of donor and acceptor chromophores with F\"orster distance
$R_0$ attached to some macromolecule, $E_t=\E{(1+[(\f x^{\rm don}_t-\f x^{\rm acc}_t)/R_0]^6)^{-1}}$,
where the simplest bound reads $T\DS\ge R_0^2\gamma(E_t-E_0)^2/8t$
(for details see \cite{Note1}).

\begin{figure}
\includegraphics[width=.475\textwidth]{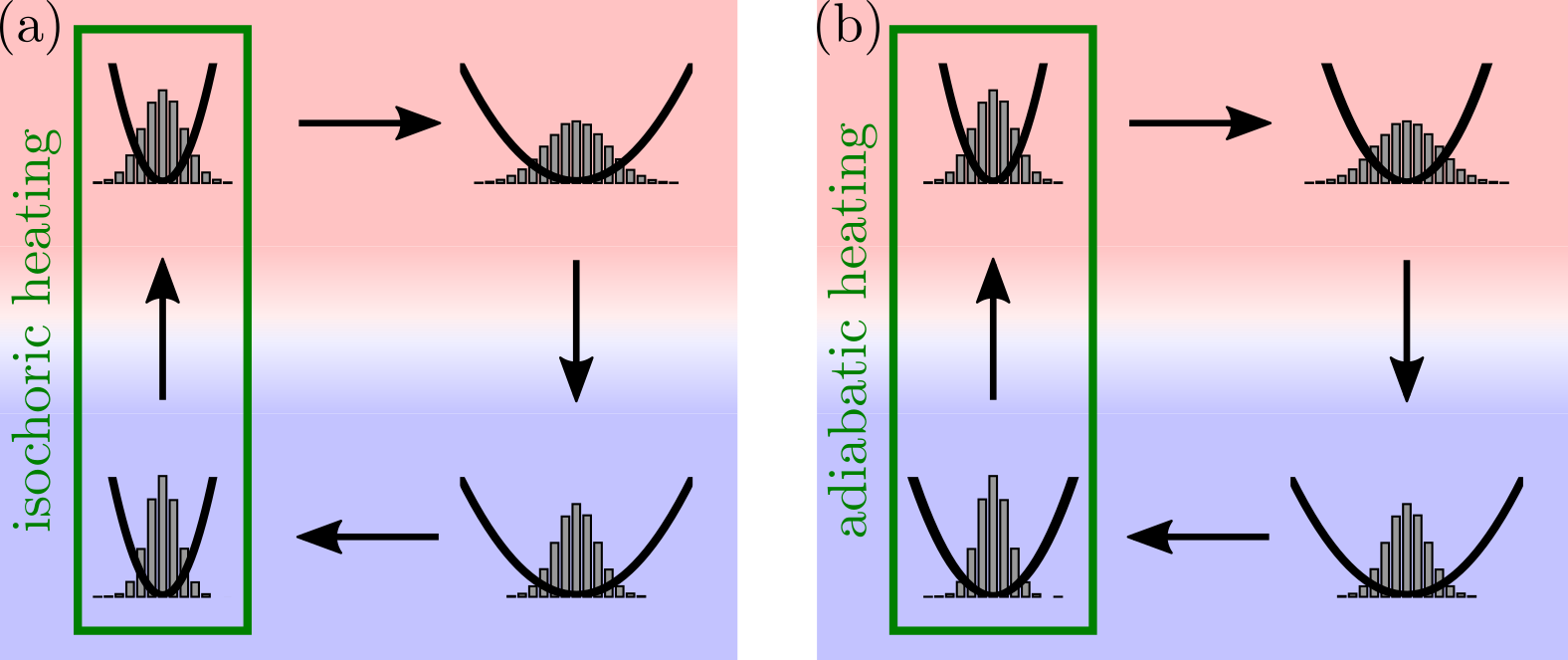}
\caption{Schematic of (a) a Stirling heat engine as realized in
  Ref.~\cite{blickle2012realization} and (b) a Carnot heat engine with adiabatic heating
  and cooling \cite{Martinez2015PRL} as
  realized in Ref.~\cite{martinez2016brownian}. Note that the stiffness is
  constant (``isochoric'') during heating and cooling in (a) but \emph{not} in
  (b). Red denotes hot and blue cold temperatures $T(\tau)$ of the
  environment (not to be confused with $T_{\rm p}$).}
\label{fig:engines}
\end{figure}

\emph{Example 3:~Higher-order transport.---}
Consider now a \emph{centered} transient process
$x_\tau$, i.e., with constant mean $\E{x_t-x_0}=0$. There is no mean
transport. However, there is higher order transport, in the simplest
case $\E{x_t^2-x_0^2}\ne 0$. Setting $z_\tau=x_\tau^2$ we then have
$T\DS\ge \gamma\E{x_t^2-x_0^2}^2/4\int_0^t\rmd\tau\E{x_\tau^2}$. A concrete example are Brownian heat engines
\cite{blickle2012realization,martinez2016brownian,Martinez2017SM} in 
Fig.~\ref{fig:engines}, where an ``effective 
temperature'' in a parabolic trap with stiffness $\kappa(\tau)$ is defined as $T_{\rm  p}(\tau)\equiv\kappa(\tau)\E{x_\tau^2}/k_{\rm B}$ \cite{Martinez2017SM}.  
In this scenario, where the medium temperature varies in
time, the left-hand side of the transport bound~\eqref{transport bound} becomes
$T\DS\to\DTS\equiv\int_0^t\rmd\tau T(\tau)\dot S_{\rm tot}(\tau)$,
yielding \footnote{Alternatively we can also derive a bound for
$\Delta S_{\rm tot}$ but we opt for $\DTS$ for a ``direct'' energetic interpretation.}
\begin{align}
\DTS\ge\frac{\left[T_{\rm p}(t)-T_{\rm p}(0)-\int_0^t T_{\rm p}(\tau)\frac{\partial_\tau\kappa(\tau)}{\kappa(\tau)}\rmd\tau\right]^2}{4\int_0^t T_{\rm p}(\tau)\kappa(\tau)\rmd\tau/k_{\rm B}\gamma}.
\label{engine bound}
\end{align}
During an isochoric heating step ($\kappa=\rm constant$ such that
  $\partial_\tau\kappa(\tau)=0$) highlighted in
Fig.~\ref{fig:engines}a, no work is performed
\cite{blickle2012realization}, and the dissipation $\DTS$ of an
efficient engine should be minimal.  Thus, to achieve a given $T_{\rm
  p}(t)-T_{\rm p}(0)$ \footnote{More precisely, a given $\gamma[T_{\rm
    p}(t)-T_{\rm p}(0)]^2/\kappa$.}
for minimal $\DTS$, \blue{assuming that we can saturate the bound (as we can in the overdamped case)}, we either need long $t$ or must maximize
$\int_0^t\rmd\tau T_{\rm p}(\tau)$, implying substantial
heating at the beginning of the time interval $[0,t]$.
For time-dependent $\kappa(\tau)$ (see Fig.~\ref{fig:engines}b or the
Carnot engine \cite{martinez2016brownian}), the bound in
Eq.~\eqref{engine bound} is more complicated since  \emph{all} terms
contribute when
$\partial_\tau\kappa(\tau)\ne0$. However, the
bound still serves as a \emph{fundamental} limit that can be evaluated for
any given protocol. Note that the results equally apply to underdamped
heat engines (as theoretically considered in, e.g.,
\cite{Dechant2017EPL}).

\blue{
Of mathematical interest for higher-order transport is the
moment-generating function $\phi_\tau(\f q)\equiv\E{{\rm e}^{-\f
    q\cdot\x_\tau}}$, giving 
\begin{align}
T\DS\ge\frac{[\phi_t(\f q)-\phi_0(\f 
q)]^2}{\int_0^t\rmd\tau\phi_\tau(2\f q)\f q^T\bgam^{-1}(\tau)\f 
q}\,,
\label{bound characteristic function} 
\end{align}
which quantifies how changes of the $q$-th mode of the probability
density contribute to $\DS$
and will be useful 
for proving future bounds.}

\emph{Interpretation and handling of $\mathcal D^z(t)$.---}The
interpretation of $\mathcal D^z(t)$ is intuitive: for a 
process $\x_\tau$ with given cost $\DS$, the transport $\E{z_t-z_0}$ will be
larger for a function $z$ that stretches $\x$ more. This stretching
is re-scaled by $\nabla_\x z$ in the quadratic
form $\mathcal{D}^z(t)$ \blue{since $\Delta S_{\rm tot}$ does \emph{not}
  depend on $z$}.

For simple marginal observations, $z(\x)=x_i$, we have that
$\mathcal{D}^z={\bgam^{-1}}_{ii}$. Moreover, if we only observe
$z_\tau$ but not $\x_\tau$, we can often bound
$\mathcal{D}^z(t)$ in terms of $\E{z_\tau}$ or in terms of
constants, such as in the case of $R_g^2$ and $S_t(q)$ before (e.g., it
suffices to know that $z_\tau$ has
bounded derivatives). In the challenging case where we only observe
$z_\tau$ but do not know the function $z(\x)$ (i.e., it is
some unknown projection or reaction coordinate), we can, given sufficient time resolution,
determine or estimate $\mathcal{D}^z$ as follows:
For overdamped dynamics we have $t\mathcal D^z(t) = 
\int_0^t\rmd\tau\frac{\var[\rmd z_\tau]}{2k_{\rm B}T\rmd\tau}$ 
(for steady-state dynamics see \cite{Dechant2023PRL}, where $\mathcal{D}^z$ recently
appeared in a correlation inequality). For
underdamped dynamics, scalar $m,\gamma$, and $z_\tau=z(\x_\tau)$ (no
explicit time dependence) we can obtain $\mathcal{D}^z(t)$ if we know the momentum relaxation
  time $m/\gamma$ via
$\frac{\var\left(\rmd\left[\frac{\rmd}{\rmd \tau}
    z(\x_\tau)\right]\right)}{2k_{\rm B}T\rmd
  t}=\frac{\gamma^2}{m^2}\E{\gamma^{-1}[\nabla_\x z(\x_\tau)]^2}$. If
  the system relaxes to equilibrium we can obtain
  $\gamma/m$ from observations of $z_\tau$ via equilibrium measurements of
  $\var(\rmd[\frac{\rmd}{\rmd \tau} z(\x_\tau)])$ and $\var(\rmd
  z(\x_\tau))$ (see \cite{Note1}). If 
  the system relaxes to a NESS, we can
  upper bound $\mathcal D^z$ 
  by using that
  $m/\gamma \le t_{\rm rel}$ with the relaxation time of the system
 determined, e.g.\ from correlation functions $t^{-1}_{\rm
    rel}=-\lim_{t\to\infty}t^{-1}\ln [\langle z_t z_0\rangle -
   \langle z_t\rangle\langle z_0\rangle]$ (see  \cite{Note1}).

 \emph{Conclusion.---}We~proved an inequality upper-bounding 
 transport of any \red{differentiable }scalar observable in a general $d$-dimensional
 non-equilibrium system in terms of the total entropy production and 
 fluctuation-scale function that ``corrects'' for the amount the
 observation stretches microscopic coordinates. We 
 explained how to
 saturate the bound. The result, classifiable as a time-integrated
generalized speed limit, may be
 understood as a thermodynamic cost of transport of observables and
 allows for inferring a lower bound on dissipation, thus
 complementing the TUR and existing speed limits.~The bound places underdamped
 and overdamped stochastic as well as deterministic systems on equal
 footing.
 This fills an important gap,
because microscopic dynamics is---at least on short time
scales---underdamped, and the TUR does not hold for
underdamped dynamics.~In particular short-time TURs
for overdamped
dynamics \cite{Manikandan2020PRL,Otsubo2020PRE,Otsubo2022CP} are
expected to fail.

 By only requiring averages, 
 the transport bound is statistically less demanding and
 applicable to both, single-molecule 
 as well as bulk experiments.~This is
 attractive in the context 
of time-resolved X-ray scattering on biomolecules, as it will allow
thermodynamic inference
from 
bulk observations of controlled transients
\cite{Dave2015CMLS,Woenckhaus2001BJ,Kubelka2009PPS,Vestergaard2007PB}.
This may facilitate
thermodynamic
inference on molecular machines without an obvious directionality
such molecular chaperones \cite{Chaperones}, which remains challenging
even with most advanced single-molecule techniques
\cite{Thorsten}. The bound allows for 
versatile applications and generalizations to vectorial 
observables $\f z$ and adaptations for Markov-jump dynamics.

\emph{Acknowledgments.---}Financial support from the European Research Council (ERC) under the European Union’s Horizon Europe research and
         innovation program (grant agreement No 101086182 to AG) is
         gratefully acknowledged.

\let\oldaddcontentsline\addcontentsline
\renewcommand{\addcontentsline}[3]{}
\bibliography{thesis_transport.bib}
\let\addcontentsline\oldaddcontentsline

\clearpage
\newpage
\onecolumngrid
\renewcommand{\thefigure}{S\arabic{figure}}
\renewcommand{\theequation}{S\arabic{equation}}
\setcounter{equation}{0}
\setcounter{figure}{0}
\setcounter{page}{1}
\setcounter{section}{0}

\begin{center}\textbf{Supplemental Material for:\\ Thermodynamic bounds on generalized transport: From single-molecule to bulk observables}\\[0.2cm]
Cai Dieball and Alja\v{z} Godec\\
\emph{Mathematical {\footnotesize bio}Physics Group, Max Planck Institute for Multidisciplinary Sciences, 37077 G\"ottingen, Germany}\\[0.6cm]\end{center}
\begin{quotation}
In this Supplemental Material we provide derivations and additional details for results and examples shown in the Letter. We provide \blue{a detailed account and comparison to existing results,} the detailed derivation of the main result, and details and parameters complementing the examples in Figs.~1 and 2. Moreover, we explain in detail how to obtain $\mathcal D^z$ from short-time fluctuations.
\end{quotation}
\renewcommand{\rmd}{{\rm d}}

\hypersetup{allcolors=black}\tableofcontents\hypersetup{allcolors=mylinkcolor}

\blue{
\newpage
\section{Comparison to existing literature}
\subsection{General discussion}
The transport bound presented in Eq.~(1) in the Letter reads
\begin{align}
&T\DS \ge \frac{\Elr{z_t-z_0-\int_0^t\partial_\tau z_\tau\rmd\tau}^2}{t\mathcal{D}^z(t)}\nonumber\\
&\mathcal{D}^z(t)\equiv\frac1t\int_0^t\E{[\nabla_\x z_\tau]^T\bgam^{-1}(\tau)\nabla_\x z_\tau}\rmd\tau,
\label{transport bound full SM} 
\end{align}
where the observable $z_\tau\equiv z(\x_\tau,\tau)$ is a scalar function of $\x_\tau$ and $\tau$. This equation is proven below for underdamped stochastic dynamics, which includes overdamped dynamics as a special case.

Taking the limit $t\to 0$ without explicit time-dependence, i.e., $z_\tau\equiv z(\x_\tau)$, Eq.~\eqref{transport bound full SM} simplified to the speed limit
\begin{align}
\left(\frac{\rmd}{\rmd\tau}\E{z_\tau}\right)^2 \le T\dot S_{\rm tot}(\tau)\E{[\nabla_\x z_\tau]^T\bgam^{-1}(\tau)\nabla_\x z_\tau}\,,
\label{transport bound as speed limit}
\end{align}
where $\dot S_{\rm tot}(\tau)\equiv\frac{\rmd}{\rmd\tau}\DS[0,\tau]$
is the entropy production rate. This result was derived for overdamped
dynamics with diffusion matrix $\f D=k_{\rm B}T\bgam^{-1}$ in
Ref.~\cite{SM_Dechant2018PRE}. However, it was neither extended to
underdamped dynamics, nor identified to apply to bulk experiments. The
importance of this extension to underdamped dynamics becomes
particularly clear in this form, since speed limits inherently address
the short-time behavior $\frac{\rmd}{\rmd\tau}\E{z_\tau}$, which (even
despite the averaging) differs substantially between for overdamped
and underdamped dynamics (see $t\to 0$ in Fig.~1b in the Letter and
divergence in Fig.~\ref{fig:SM_TUR_underdamped}). Also note that in
the underdamped setting (colored lines in Fig.~1b in the Letter), the
maximal quality of the bound is not at $t\to 0$, which (besides being
more directly applicable) is a strong argument for preferring the time-integrated version over the speed limit (the latter corresponds to $t\to 0$).

Further restricting to one-dimensional space and $z_\tau=x_\tau$, i.e., to the bound 
\begin{align}
\left (\frac{\rmd}{\rmd\tau}\E{x_\tau}\right )^2 \le D\dot S_{\rm tot}(\tau)/k_{\rm B}.
\label{simplest transport bound as speed limit} 
\end{align}
This transport bound corresponds in periodic systems (see also following section) to speed limits from Refs.~\cite{SM_Leighton2022PRL,SM_Leighton2023PRL} which were applied to gain important theoretical insights into the efficiency and energetics of molecular motors. 

The bound~\eqref{transport bound full SM} complements the Benamou-Brenier formula \cite{SM_Benamou2000NM} that applies to overdamped Markov observables \cite{SM_VanVu2023PRX,SM_Dechant2022JPA} by
allowing for the more realistic underdamped dynamics, general projected (non-Markovian) observables $z_\tau$, and time-dependent diffusivity. While the Benamou-Brenier formula gives a lower bound on the entropy production from the difference in initial and final probability densities, we directly target more easily measurable observables.

\subsection{Spatially periodic systems}
Systems that are periodic in space (such as, e.g., Brownian motion on a ring or an interval with periodic boundary conditions) are paradigmatic examples for non-equilibrium steady states \cite{SM_Seifert2012RPP}, in particular to model molecular motors \cite{SM_Seifert2018PA,SM_Leighton2022PRL}. Such systems can either be viewed as transient (e.g., with angle on the ring in $(-\infty,\infty)]$) or as steady states (e.g., angle in $[0,2\pi)$). This complementary picture allows to apply both, the transport bound and the thermodynamic uncertainty relation (TUR) \cite{SM_Barato2015PRL}, to such systems.

In Fig.~\ref{fig:SM_TUR_transport_bootstrap} we evaluate for simulated overdamped Langevin motion
\begin{align}
\rmd x_\tau = [D\partial_x \ln\ps(x_\tau)+\nu_{\rm s}(x_\tau)]\rmd\tau+\sqrt{2D}\rmd W_\tau\,,\label{LE periodic} 
\end{align}
with $D=k_{\rm B}T/\gamma$ and periodic functions $\ps$ and $\nu_{\rm s}$  different versions of the TUR and transport bound, given by the equations (see Refs.~\cite{SM_Dechant2021PRX, SM_Dieball2023PRL} for correlation-TUR (CTUR))
\begin{align}
J_t^x &\equiv \int_{\tau=0}^{\tau=t}1\circ\rmd x_\tau=x_t-x_0\nonumber\\
J_t^\nu &\equiv \int_{\tau=0}^{\tau=t}\nu(x_\tau)\circ\rmd x_\tau\nonumber\\
\rho_t^\nu &\equiv \int_0^t\nu^2(x_\tau)\rmd\tau\nonumber\\
z^\nu_\tau &\equiv \int_0^{x_\tau}\nu(x)\rmd x\nonumber\\
Q_{{\rm TUR}\, x/\nu} &\equiv \frac{2\E{J_t^{x/\nu}}^2}{\var(J_t^{x/\nu})}\,\frac{k_{\rm B}}{\DS}\nonumber\\
Q_{{\rm CTUR}\, \nu} &\equiv \frac{2\E{J_t^\nu}^2}{\var(J_t^\nu)\left[1-\frac{\cov(J_t^\nu,\rho_t^\nu)^2}{\var(J_t^\nu)\var(\rho_t^\nu)}\right]}\,\frac{k_{\rm B}}{\DS}\nonumber\\
Q_{{\rm transport}\, x} &\equiv \frac{{\E{x_t-x_0}^2}}{Dt}\,\frac{k_B}{\DS}\nonumber\\
Q_{{\rm transport}\, \nu} &\equiv \frac{\E{z^\nu_t-z^\nu_0}^2}{D\E{(\partial_x z^\nu_\tau)^2}}\,\frac{k_B}{\DS}\,.
\label{periodic transport TUR}
\end{align}
\begin{figure}
\includegraphics[scale=1]{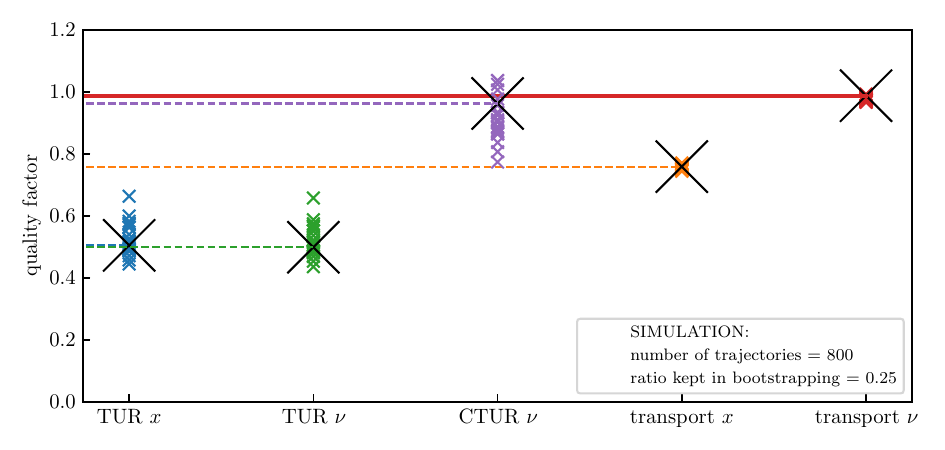}
\caption{\blue{Evaluation of the bounds in Eq.~\eqref{periodic transport TUR} for 800 trajectories (big black crosses) for the Langevin motion~\eqref{LE periodic} with $k_{\rm B}T=1$, $t=1800$, $D=0.01$, $\ps(x)=f(x)/\int_0^L f(x')\rmd x'$ with $f(x)=1.7 + \cos(2\pi x/L) + 0.3\sin(4\pi x/L) + 0.2\cos(8\pi x/L)$ and periodicity $L=5$, $\nu_{\rm s}(x)=0.01/\ps(x)$ simulated with Euler time step $\rmd t=0.04$. In practical applications, most parameters are not known. Therefore, for the evaluation of Eq.~\eqref{periodic transport TUR}, $\nu_{\rm s}(x)$ is inferred from the trajectories by inferring $\ps$ from a histogram projected on $[0,L]$ and subsequently fitting the product $\nu_{\rm s}(x)\ps(x)$ which must be constant in one-dimensional steady-state dynamics \cite{SM_Gardiner1985}. $D$ is obtained from the (overdamped) short time fluctuations $D=\var(\rmd x_\tau)/2\rmd\tau$ averaged over all times $\tau\in[0,t]$. To address the fluctuations of the result, a bootstrapping procedure has been employed where 25 times the quality factor is derived from a random subset of 25 \% of the trajectories (colored crosses).}}
\label{fig:SM_TUR_transport_bootstrap}
\end{figure}
For this example, the transport bound $Q_{{\rm transport}\, x}$
performs better than the TUR. However, there are also overdamped
periodic examples where the TUR performs performs better
\cite{SM_Leighton2022PRL}. As expected, the absence of variances in the
transport bound leads to a statistically more robust inference as can
be seen from the spread of the colored crosses in
Fig.~\ref{fig:SM_TUR_transport_bootstrap} (i.e.\ the statistical noise
due to undersampling is substantially smaller when using the transport
bound). $Q_{{\rm TUR}\, \nu}$ was in included in the comparison since it is known to approach $1$ close to equilibrium \cite{SM_Pigolotti2017PRL} and for short times \cite{SM_Manikandan2020PRL}. However, it does not exceed $Q_{{\rm TUR}\, x}$ in this example.

The analytical value for the CTUR with the optimal choice of
$J_t^\nu,\,\rho_t^\nu$ is $Q_{{\rm CTUR}\,\nu}=1$
\cite{SM_Dechant2021PRX,SM_Dieball2023PRL}. This is roughly achieved for 800
long trajectories. However, with ``only'' 200 trajectories the result still fluctuates a lot. The saturated transport bound with the choice of $z_\tau^\nu$ is also very close to the analytical value of $1$ and fluctuates much less, i.e., this bound is accessible from less trajectories as compared to the TUR.

Compared to the TUR, the transport bound has the additional advantage of being valid for transients and underdamped motion, i.e., the inference will not fail in case these underlying assumptions are not justified (one only has to be careful what assumptions one wants to employ for evaluating $\mathcal D^z$). For an illustration of the failure of the overdamped TUR see Fig.~\ref{fig:SM_TUR_underdamped}. in Sec.~\ref{subsec:TUR underdamped} (transient example; see Ref.~\cite{SM_Patrick_break} for a steady-state example; see Supplemental Material of Ref.~\cite{SM_Dieball2023PRL} for an example where the overdamped steady-state TUR fails for overdamped transients).}

\section{Proof of the main result}
Here, we prove our main result, i.e., the transport bound in Eq.~(1) in the Letter. 
\blue{We first give a short sketch \red{(which only serves to give an optional overview and may be skipped)} and afterwards write the full proof. Finally, we will also show  how the proof dramatically simplifies in the special case of overdamped dynamics.

\subsection{Sketch of proof}
To prove Eq.~(1) in the Letter \red{for an observable $z_\tau$}, we use the Cauchy-Schwarz
inequality for stochastic integrals similarly as in
\cite{SM_Dieball2023PRL} but generalized to underdamped dynamics. 
\red{Recalling $\bgam$ and $\rmd\f W_\tau$ and the $\bnu$ from Eqs.~(2) and (4) in the Letter, respectively, we} define the stochastic integrals
$A_t\equiv\int_{\tau=0}^{\tau=t}\bnu(\x_\tau,\f
v_\tau,\tau)\cdot\sqrt{\bgam}\rmd\f W_\tau$ and
$B_t\equiv\int_{\tau=0}^{\tau=t}\{\nabla_\x
z_\tau\}\cdot{\sqrt{\bgam}^{-1}}^{T}\rmd\f W_\tau$ for which we can
show that $\E{A_t^2}=T\DS$ and $\E{B_t^2}=t\mathcal D^z(t)$\red{, where $\DS$ and $D^z(t)$ are defined in Eqs.~(5) and (1) in the Letter, respectively}. The Cauchy-Schwarz inequality thus implies $T\DS\ge \E{A_tB_t}^2/t\mathcal{D}^z(t)$. To complete the proof, we compute $\E{A_tB_t}=\E{\int_0^t\bnu(\x_\tau,\f v_\tau,\tau)\cdot\left\{\nabla_\x z_\tau\right\}\rmd\tau}$ via rewriting $\{\nabla_\x z_\tau\}=\bgam\f m^{-1}\nabla_{\f v}\{\nabla_{\f v}^T\f m\bgam^{-1}\nabla_\x z_\tau\}$ \red{[with $\f m$ from Eq.~(2) in the Letter]}, integrating by parts in $\nabla_{\f v}$, and substituting the Klein-Kramers equation \red{(see Eq.~\eqref{fourth attempt eom} below)} in the form $\nabla_\f v\cdot\f m^{-1}\bgam\bnu\rho=[\nabla_{\f x}\cdot\f v+\nabla_{\f v}\cdot\f m^{-1}\f F(\x,\tau)+\partial_\tau]\rho(\x,\f v,\tau)$,
which upon further integrations by parts and simplifications yields
$\E{A_tB_t}=\int_0^t\rmd\tau\Elr{\f v_\tau\cdot\nabla_\x
  z_\tau}=\E{\int_0^t\rmd\tau(\frac{\rmd}{\rmd
    \tau}-\partial_\tau)z_\tau}=\E{z_t-z_0-\int_0^t\partial_\tau
  z_\tau\rmd\tau}$\red{, completing the proof of Eq.~(1) in the Letter.}

From the proof we immediately know how to achieve
saturation by saturating the Cauchy-Schwarz inequality, allowing
optimal inference. Saturation occurs for $\nabla_\x
z_\tau=c\bgam\bnu(\x,\f v,\tau)$ for any constant $c$. While one may
not always be able to choose $z$ this way, one should aim to approach this for saturation.
If $d=1$ (with
natural boundaries) and  $\bnu$ does not depend on $\f v$,
saturation is always feasible.}

\subsection{Full proof}
Consider underdamped stochastic dynamics as in Eq.~(2) in the Letter. Consider positive definite symmetric matrices $\f m$ and $\bgam(\tau)$ (for simplicity denote the latter by $\bgam$). Using the ``local mean velocity'' $\bnu$ as in Eq.~(4) in the Letter we can write the underdamped Klein-Kramers dynamics as \cite{SM_Risken1989}
\begin{align}
\bnu(\x,\f v,\tau)&\equiv\f v_\tau+k_{\rm B}T\f m^{-1}\frac{\nabla_{\f v}\rho(\x,\f v,t)}{\rho(\x,\f v,t)}\nonumber\\
\partial_\tau \rho(\x,\f v,t) &= \left(-\nabla_\x\cdot\f v+\f m^{-1}\nabla_{\f v}\cdot\left[-\f F(\x,\tau)+{\bgam(\tau)\bnu(\x,\f v,\tau)}\right]\right)\rho(\x,\f v,t)\,.\label{fourth attempt eom} 
\end{align}
Recalling the setup of the proof as presented above, and using the white noise property/Wiener correlation $\E{\rmd W_\tau^i\rmd W_{\tau'}^j}=\delta(\tau-\tau')\rmd\tau\rmd\tau'$ we have
\begin{align}
A_t&\equiv\int_{\tau=0}^{\tau=t}\bnu(\x_\tau,\f v_\tau,\tau)\cdot\sqrt{\bgam}\rmd\f W_\tau\nonumber\\
B_t&\equiv\int_{\tau=0}^{\tau=t}\left\{\nabla_\x z(\x_\tau,\tau)\right\}\cdot{\sqrt{\bgam}^{-1}}^{T}\rmd\f W_\tau\nonumber\\
\E{A_t^2}&=\int_0^t\rmd\tau\E{\bnu^T(\x_\tau,\f v_\tau,\tau)\bgam\bnu(\x_\tau,\f v_\tau,\tau)}=T\DS([0,t])\nonumber\\
\E{B_t^2}&=\int_0^t\rmd\tau\Elr{\left\{\nabla_\x z(\x_\tau,\tau)\right\}^T\bgam^{-1}\left\{\nabla_\x z(\x_\tau,\tau)\right\}}\equiv t\mathcal D^z(t)\,.\label{fourth attempt A2B2} 
\end{align}
Using the Einstein summation convention, we can check the matrix
differential operator identity for a constant (i.e.\ $\x$-independent) invertible $d\times d$ 
matrix $\f G$
\begin{align}
(\f G\nabla_{\f v})[\f v\cdot\f G^{-1}\nabla_\x z(\x,\tau)]=\f e_iG_{ij}\partial_{v_j}v_n(\f G^{-1})_{nm}\partial_{x_m} z(\x,\tau)=\f e_iG_{in}(\f G^{-1})_{nm}\partial_{x_m} z(\x,\tau)=\nabla_\x z(\x,\tau)\,.
\end{align}
Using this identity for the choice $\f G^{-1}=\f m\bgam^{-1}$, integrating by parts, and substituting Eq.~\eqref{fourth attempt eom} for $\nabla_{\f v}\cdot\f m^{-1}\bgam\bnu$, we obtain
\begin{align}
-\E{A_tB_t}&=-\int_0^t\rmd\tau\int\rmd\x\int\rmd\f v\left\{\f G\nabla_{\f v} [\f v\cdot\f G^{-1}\nabla_\x z(\x,\tau)]\right\}\cdot\bnu(\x,\f v,\tau)\rho(\x,\f v,\tau)\nonumber\\
&=\int_0^t\rmd(\x,\f v,\tau)\{\f v\cdot
\f G^{-1}\nabla_\x z(\x,\tau)\}\nabla_{\f v}\cdot\f G^T\bnu(\x,\f v,\tau)\rho(\x,\f v,\tau)
\nonumber\\
&=\int_0^t\rmd(\x,\f v,\tau)\{\f v\cdot
\f G^{-1}\nabla_\x z(\x,\tau)\}\f m^{-1}\nabla_{\f v}\cdot\bgam\bnu(\x,\f v,\tau)\rho(\x,\f v,\tau)
\nonumber\\
&=\int_0^t\rmd\tau\int\rmd\x\int\rmd\f v\{\f v\cdot\f G^{-1}\nabla_\x z(\x,\tau)\}\left[\nabla_{\f x}\cdot\f v+\f m^{-1}\nabla_{\f v}\cdot\f F(\x,\tau)+\partial_\tau\right]\rho(\x,\f v,\tau)\nonumber\\
&=\Elr{\f v_t\cdot\f G^{-1}\nabla_\x z(\x_t,t)-\f v_0\cdot\f G^{-1}\nabla_\x z(\x_0,0)}\label{proof calculation}\\
&\quad 
-\f m\int_0^t\rmd\tau\Elr{(\f v_\tau\cdot\nabla_\x)[\f v_\tau\cdot\f G^{-1}\nabla_\x z(\x_\tau,\tau)]+\f m^{-1}\f F(\x_\tau,\tau))\cdot\f G^{-1}\nabla_\x z(\x_\tau,\tau)+\f v_\tau\cdot\f G^{-1}\nabla_\x \partial_\tau z(\x_\tau,\tau)}\,\nonumber.
\end{align}
Note that 
\begin{align}
\Elr{\f v_t\cdot\f G^{-1}\nabla_\x z(\x_t,t)-\f v_0\cdot\f G^{-1}\nabla_\x z(\x_0,0)}=\Elr{\int_0^t\rmd\tau\frac{\rmd}{\rmd \tau}[\f v_\tau\cdot\f G^{-1}\nabla_\x z(\x_\tau,\tau)]}\,,
\end{align}
and use the equation of motion in Eq.~\eqref{fourth attempt eom} to see that
\begin{align}
\Elr{\frac{\rmd}{\rmd \tau}[\f v_\tau\cdot\f G^{-1}\nabla_\x z(\x_\tau,\tau)]}&=\Elr{\f m^{-1}\left(\f F(\x_\tau,\tau)-\bgam\f v_\tau\right)\cdot\f G^{-1}\nabla_\x z(\x_\tau,\tau)+\f v_\tau\cdot\f G^{-1}\nabla_\x[\f v_\tau\cdot\nabla_\x z(\x_\tau,\tau)+\partial_\tau z(\x_\tau,\tau)]}\,.
\end{align}
Since $\f G$ is independent of $\x$ we have 
\begin{align}
(\f v_\tau\cdot\nabla_\x)[\f v_\tau\cdot\f G^{-1}\nabla_\x z(\x_\tau,\tau)]=(\f v_\tau\cdot\f G^{-1}\nabla_\x)[\f v_\tau\cdot\nabla_\x z(\x_\tau,\tau)]\,,
\end{align}
which allows to conclude that
\begin{align}
-\E{A_tB_t}&=-\int_0^t\rmd\tau\Elr{\f v_\tau\cdot\nabla_\x z(\x_\tau,\tau)}\,.
\end{align}
Moreover, using
\begin{align}
\frac{\rmd}{\rmd \tau}z(\x_\tau,\tau)=\f v_\tau\cdot\nabla_\x z(\x_\tau,\tau)+\partial_\tau z(\x_\tau,\tau)\,,
\end{align}
we obtain
\begin{align}
\E{A_tB_t}&=\Elr{z(\x_t,t)-z(\x_0,0)-\int_0^t\rmd\tau\partial_\tau z(\x_\tau,\tau)}\,.\label{proof end calculation}
\end{align}
With Eq.~\eqref{fourth attempt A2B2} this proves via Cauchy-Schwarz $\E{A_tB_t}^2\le \E{A_t^2}\E{B_t^2}$ the transport bound
\begin{align}
\frac{\Elr{z(\x_t,t)-z(\x_0,0)-\int_0^t\rmd\tau\partial_\tau z(\x_\tau,\tau)}^2}{t\mathcal D^z(t)/k_{\rm B}T}
\le T\DS\,.\label{transport bound SM} 
\end{align}

\subsection{Overdamped limit}
Although the underdamped results already contain the overdamped and Newtonian cases as limiting results, it is interesting to see that the derivation in the overdamped case greatly simplifies. Here, we have that $\nabla_\x\cdot\bnu(\x,\tau)\rho(\x,\tau)=-\partial_\tau\rho(\x,\tau)$ \cite{SM_Risken1989} such that Eq.~\eqref{transport bound SM} directly follows from
\begin{align}
\E{A_tB_t}&=\int_0^t\rmd\tau\int\rmd\x\left\{\nabla_\x z(\x,\tau)]\right\}\cdot\bnu(\x,\tau)\rho(\x,\tau)\nonumber\\
&=\int_0^t\rmd\tau\int\rmd\x z(\x,\tau)\partial_\tau\rho(\x,\tau)\nonumber\\
&=\Elr{z(\x_t,t)-z(\x_0,0)-\int_0^t\rmd\tau\partial_\tau z(\x_\tau,\tau)}
\,.\label{derivation overdamped} 
\end{align}

\blue{\subsection{Generalization to velocity dependent variables}
For optimal inference in the case of complete observation of $(\x_\tau,\f v_\tau)$ it is also useful to generalize Eq.~\eqref{transport bound full SM} to velocity-dependent observables $z(\x_\tau,\f v_\tau,\tau)$. We now show how one can generalize the proof starting with $-\E{A_tB_t}=-\int_0^t\rmd\tau\int\rmd\x\int\rmd\f v\left\{\nabla_\x z(\x,\f v,\tau)\right\}\cdot\bnu(\x,\f v,\tau)\rho(\x,\f v,\tau)$ as in Eq.~\eqref{proof calculation} where $\bnu(\x,\f v,\tau)\equiv\f v_\tau+k_{\rm B}T\f m^{-1}\frac{\nabla_{\f v}\rho(\x,\f v,t)}{\rho(\x,\f v,t)}$. While the first term ($\f v_\tau$ in $\bnu(\x,\f v,\tau)$) will similarly as in Eqs.~\eqref{proof calculation}-\eqref{proof end calculation} yield $\E{z(\x_t,\f v_t,t)-z(\x_0,\f v_0,0)-\int_0^t\rmd\tau\left[\partial_\tau z(\x_\tau,\f v_\tau+\dot{\f v}\partial_{\f v} z(\x_\tau,\f v_\tau,\tau)\right]}^2$. While $\dot{\f v}$ contains the white noise part that takes very high values, its occurrence in the average value should not be problematic for sufficient statistics since the high values average out for any finite time resolution. Therefore this expression will remain operationally accessible if enough trajectories are measured and one freely chooses $z_\tau$ to optimize inference.

The second term of $\bnu$ contributes in the presence of $\f v$-dependence in $z_\tau$ as 
\begin{align}
&k_{\rm B}T\int_0^t\rmd\tau\int\rmd\x\int\rmd\f v\left\{\nabla_\x z(\x,\f v,\tau)\right\}\cdot \f m^{-1}\nabla_{\f v}\rho(\x,\f v,t)\nonumber\\
&\qquad=-k_{\rm B}T\int_0^t\rmd\tau\int\rmd\x\int\rmd\f v\rho(\x,\f v,t)\nabla_{\f v}\cdot \f m^{-1}\nabla_\x z(\x,\f v,\tau)\,.\label{new term v dependence}
\end{align}
This term did not contribute before since we had $\nabla_{\f v}z_\tau=0$. However, this term can also be obtained from trajectories by computing the operationally accessible integral $\int\rmd(\nabla_\x z)\cdot\rmd\f v$ where only the noise part contributes giving rise to the term in Eq.~\eqref{new term v dependence} (see Ref.~\cite{SM_arxiv_Crutchfield_TUR}).

Following these lines, one can generalize and apply the transport bound even to velocity dependent observables. Note however, that the result becomes slightly more complicated, and that much of the intuition on spatial transport under an external friction is lost in the case of explicit velocity-dependence.}

\section{Details on the example in Fig.~1 in the Letter}
Here, we give details on the example of the displaced trap considered
in Fig.~1 in the Letter. Note that the overdamped case of this example
including application of the TUR can be found in
\cite{SM_Dieball2023PRL}. In Figure~1 we consider the parameters $a=5$, $m=1$ and $y=1$. In the overdamped case, $\gamma$ cancels out in the quality factor.

The underdamped equations of motion for this example for a linear
force originating from a harmonic potential around $0$ with a stiffness $a$ read 
\begin{align}
\rmd\begin{bmatrix}x_\tau\\v_\tau\end{bmatrix}&=-\underbrace{\begin{bmatrix} 0& -1\\ a\frac\gamma m & \frac\gamma m \end{bmatrix}}_{\equiv\f A}\begin{bmatrix}x_\tau\\v_\tau\end{bmatrix}\rmd\tau+\underbrace{\frac{\sqrt{2k_{\rm B}T\gamma}}m\begin{bmatrix}0& 0\\ 0& 1\end{bmatrix}}_{\equiv\bsig}\rmd\f W_\tau\nonumber\\
\f D&=\frac{\bsig\bsig^T}2=\frac{k_{\rm B}T\gamma}{m^2}\begin{bmatrix}0& 0\\ 0& 1\end{bmatrix}\,.
\end{align}
The steady-state covariance matrix $\bSig_{\rm s}$ and time-dependent covariance $\bSig(t)$ obey the Lyapunov equations (see, e.g., Supplemental Material of Ref.~\cite{SM_Dieball2023PRR} for a short derivation)
\begin{align}
\f A\bSig_{\rm s}+\bSig_{\rm s}\f A^T&=2\f D=\bsig\bsig^T \nonumber\\
\bSig(t)&=\bSig_{\rm s}+\rme^{-\f At}[\bSig(0)-\bSig_{\rm s}]\rme^{-\f A^Tt}\,.\label{Lyapunov} 
\end{align}
We assume the particle to be equilibrated in the trap around $\E{x_0}=y$ before displacing the trap at time $t=0$ to position $x=0$. Therefore we have
\begin{align}
\bSig(0)=\bSig_{\rm s}&=k_{\rm B}T\begin{bmatrix} \frac1{a\gamma}& 0\\ 0 & \frac1m\end{bmatrix}\,,
\end{align}
which can be also obtained from equipartition. 

We see from Eq.~\eqref{Lyapunov} that $\bSig(\tau)=\bSig_{\rm s}$ for all $\tau\ge0$, i.e., the covariance remains unchanged, and in particular there will be no coupling of positions and velocities, and the density reads
\begin{align}
\rho(x,v,\tau)=\frac{\sqrt{a\gamma m}}{2\pi}\exp\left[-\frac{a\gamma(x-\E{x_\tau})^2+m(v-\E{v_\tau})^2}{2k_{\rm B}T}\right]\,.
\end{align}

The mean starting from $\E{x_0}=y$ and $\E{v_0}=0$ is governed by
\begin{align}
\begin{bmatrix} \E{x_\tau} \\ \E{v_\tau}\end{bmatrix}=\exp(-\f A\tau)\begin{bmatrix} y \\ 0 \end{bmatrix}\,.
\end{align}

The decisive term for the entropy production is [see Eqs.~(4) and (5) in the Letter]
\begin{align}
T\DS&=\gamma\int_0^t\rmd\tau\Elr{\left[v_\tau+\frac{k_{\rm B}T}{m}\,\frac{\partial_v\rho(x_\tau,v_\tau,\tau)}{\rho(x_\tau,v_\tau,\tau)}\right]^2}\nonumber\\
&=\gamma\int_0^t\rmd\tau\Elr{\left[v_\tau-\frac{k_{\rm B}T}{m}\,\frac{2m(v_\tau-\E{v_\tau})}{2k_{\rm B}T}\right]^2}\nonumber\\
&=\gamma\int_0^t\rmd\tau\Elr{v_\tau}^2\,.\label{displaced trap entropy} 
\end{align}
Note that in this simple example, the local mean velocity is not
local, i.e., it does not depend on $x_\tau$. In the overdamped case we
have $\nu(\tau)=-az\rme^{-a\tau}$ (see Supplemental Material of
Ref.~\cite{SM_Dieball2023PRL}), but in the underdamped case the
eigenvalues of $\f A$ can become complex (signaling oscillations)
\begin{align}
\f A&=\begin{bmatrix}0&-1\\a\frac\gamma m&\frac\gamma m\end{bmatrix}\nonumber\\
0&=\lambda^2-\frac\gamma m \lambda+a\frac\gamma m\nonumber\\
\lambda_{1,2}&=\frac\gamma{2m}\left[1\pm\sqrt{1-\frac{4am}{\gamma}}\,\right]\,.
\end{align}
We obtain oscillations for $4am>\gamma$, i.e., for weak damping. 

\subsection{Saturation}

We know from the Cauchy-Schwarz proof of Eq.~\eqref{transport bound SM} that we obtain saturation for $\gamma\nu(x,v,\tau)=c\partial_x z(x,\tau)$, i.e., in this case the optimal observable $z_{\rm opt}$ is given by
\begin{align}
\gamma\nu(x,v,\tau)&=\gamma\nu(\tau)\overset{\rm as\ in\ Eq.~\eqref{displaced trap entropy}}=\gamma \E{v_\tau}\equiv \gamma f(\tau)
\nonumber\\
z_{\rm opt}(x,\tau)&=\gamma x\E{v_\tau}=\gamma x\underbrace{\left(\begin{bmatrix} 0 \\ 1 \end{bmatrix}\cdot\exp(-\f A\tau)\begin{bmatrix} y \\ 0 \end{bmatrix}\right)}_{=f(\tau)}.
\end{align}
Since the Cauchy-Schwarz inequality becomes saturated for this choice, we know that the transport bound Eq.~\eqref{transport bound SM} is saturated, allowing optimal inference of $\DS$. 

This predicted saturation can be checked for this case by computing 
\begin{align}
T\DS&=\gamma\int_0^t\rmd\tau f^2(\tau)
\nonumber\\
t\mathcal D^{z_{\rm opt}}(t)&=\gamma\int_0^t\rmd\tau f^2(\tau)=T\DS
\nonumber\\
\Elr{z_{\rm opt}(\x_t,t)-z_{\rm opt}(\x_0,0)-\int_0^t\rmd\tau\partial_\tau z_{\rm opt}(\x_\tau,\tau)}
&=\gamma\E{x_t} f(t)-\gamma\E{x_0}\underbrace{f(0)}_{=0}-\gamma\int_0^t\rmd\tau\E{x_\tau}\partial_\tau f(\tau)=
\nonumber\\
&=\gamma\int_0^t\rmd\tau f(\tau)\partial_\tau\E{x_\tau}=\gamma\int_0^t\rmd\tau f^2(\tau)=T\DS\,.
\end{align}
Hence we indeed have saturation since the transport bound \eqref{transport bound SM} here reads $\DS^2/\DS\overset{\rm here\ =}\le\DS$.

\subsection{Short-time limit}
In Fig.~1b we see that the short-time limit of the quality factor is $3/4$. This can be confirmed analytically as follows,
\begin{align}
\begin{bmatrix}\E{x_\tau} \\ \E{v_\tau}\end{bmatrix}=\exp(-\f A\tau)\begin{bmatrix} y \\ 0 \end{bmatrix}=\begin{bmatrix} y \\ 0 \end{bmatrix}-\tau\begin{bmatrix}0&-1\\a\frac\gamma m & \frac\gamma m\end{bmatrix}\begin{bmatrix} y \\ 0 \end{bmatrix}+\mathcal O(\tau^2)
=\begin{bmatrix} y \\ -\frac{\tau az\gamma}m \end{bmatrix}+\frac{\tau^2}2\left[\begin{matrix}- \frac{\gamma a y}{m}\\\frac{\gamma^{2} a y}{m^{2}}\end{matrix}\right]+\mathcal O(\tau^3)\,.
\end{align}
Thus, the short-time entropy production reads
\begin{align}
T\DS &= \frac{\gamma^3t^3a^2y^2}{3m^2}+\mathcal O(t^4)\,.
\end{align}
The short-time transport reads
\begin{align}
\textnormal{bound} &= \frac{\gamma}{t}(\E{x_t}-y)^2=\frac{\gamma}{t}\frac{t^4}4\frac{\gamma^2a^2y^2}{m^2}+\mathcal O(t^4)=\frac34T\DS+\mathcal O(t^4)\,,\label{quality trap short time limit} 
\end{align}
such that the short-time quality factor is indeed $3/4$.

\blue{
\subsection{TUR applied to underdamped dynamics}\label{subsec:TUR underdamped}
The TUR \cite{SM_Barato2015PRL}, and in particular the transient version
\cite{SM_Dechant2018JSMTE}, only applies to overdamped systems. Though
there are versions for underdamped dynamics these are not practical to
use and in particular not operationally accessible, since, e.g., the
bound does not directly follow from measured trajectories but instead
requires variations of many different parameters in the system
\cite{SM_Kwon2022NJP,SM_Lee2023PRR}. However, the overdamped transient TUR
with corresponding quality factor (for simplicity we take the special case $J_t=x_t-x_0$)
\begin{align}
Q_{\rm TUR\, transient} = \frac{2(t\partial_t\E{x_t-x_0})^2}{\var(x_t-x_0)}\,\frac{k_{\rm B}}{\DS}\,,\label{quality TUR transient}
\end{align}
fails for underdamped dynamics. 

To illustrate that, we again consider the example of the displaced trap and compute $\var(x_t-x_0)=\var(x_t^2)+\var(x_0^2)+2\cov(x_t,x_0)=2[\frac{k_{\rm B}T}{a\gamma}-\cov(x_t,x_0)]$, which similarly to the above follows from 
\begin{align}
\rmd \Elr{\begin{bmatrix}x_\tau\\v_\tau\end{bmatrix}\begin{bmatrix}x_0 & v_0\end{bmatrix}} &= -\f A\Elr{\begin{bmatrix}x_\tau\\v_\tau\end{bmatrix}\begin{bmatrix}x_0 & v_0\end{bmatrix}}\rmd\tau\nonumber\\
\Elr{\begin{bmatrix}x_\tau\\v_\tau\end{bmatrix}\begin{bmatrix}x_0 & v_0\end{bmatrix}} - \Elr{\begin{bmatrix}x_\tau\\v_\tau\end{bmatrix}}\Elr{\begin{bmatrix}x_0 & v_0\end{bmatrix}} &= \rme^{-\f A\tau}\bSig_{\rm s}
\nonumber\\
\cov(x_\tau,x_0)&= \left(\rme^{-\f A\tau}\bSig_{\rm s}\right)_{xx}\nonumber\\
\var(x_t-x_0) &= 2\left[\left(\mathbf 1-\rme^{-\f At}\right)\bSig_{\rm s}\right]_{xx}\,.
\end{align}
For Eq.~\eqref{quality TUR transient}, we also require the expression
\begin{align}
\partial_t\E{x_t-x_0} &= \partial_t\E{x_t} 
= \partial_t\left(\begin{bmatrix} 1 \\ 0 \end{bmatrix}\cdot\exp(-\f At)\begin{bmatrix} y \\ 0 \end{bmatrix}\right)
 = -\left(\begin{bmatrix} 1 \\ 0 \end{bmatrix}\cdot\exp(-\f At)\f A\begin{bmatrix} y \\ 0 \end{bmatrix}\right)\,.
\end{align}
Using these results, we plot the quality factor~\eqref{quality TUR transient} for the underdamped dynamics displaced trap in Fig.~\ref{fig:SM_TUR_underdamped}a. We see that the bound $Q\le 1$ heavily violated, i.e., that the overdamped transient TUR (unlike the transport bound) is not valid for underdamped dynamics. This is partially explained due to the different time-evolution of the total entropy production, see  Fig.~\ref{fig:SM_TUR_underdamped}b. However, for small friction the bound $Q\le 1$ for the transient TUR is even violated for longer times where $\DS$ almost agrees which is due to the different character of the dynamics. 
\begin{figure}
    \centering
    \includegraphics{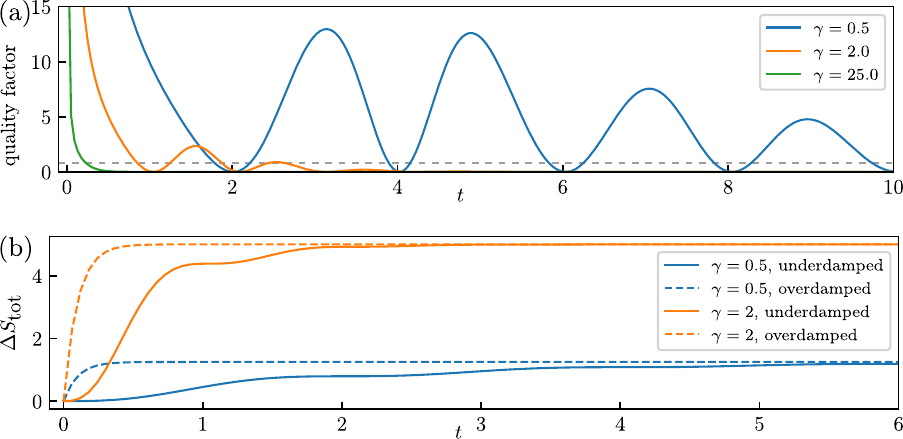}
    \caption{\blue{(a) Quality factor in Eq.~\eqref{quality TUR transient} evaluated for underdamped dynamics with different friction (same parameters as in Fig.~1 in the Letter). Values above $1$ (dashed gray line) show that the  overdamped TUR bound is violated for this underdamped example. (b) Time-evolution of the entropy production up to time $t$. As expected, underdamped and overdamped dynamics give rise to very different short-time behavior.)}}
    \label{fig:SM_TUR_underdamped}
\end{figure}
}

\section{Derivation of Eq.~(6) in the Letter}
Here, we apply and rearrange the transport bound for the radius of gyration and structure factor, reproducing the bounds presented in Eq.~(6) in the Letter.

\subsection{Radius of gyration}
First, to approach the radius of gyration consider the following observable $z$, which yields $NR_g^2=\E{z_\tau}$,
\begin{align}
\vec c &\equiv\frac1N\sum_{k=1}^N \vec x_k\nonumber\\
z(\x)&\equiv\sum_{j=1}^N(\vec x_j-\vec c\,)^2\nonumber\\
NR_g^2(t)&=\E{z(\x_t)}\,,
\end{align}
where the $\vec x_j$ are the 3d position vectors of the individual beads. Here, we denote 3d vectors by arrows and boldface notation is only used for the $3N$ dimensional vectors that contain all bead positions.

To compute $\mathcal{D}^z$ we need to compute derivatives, i.e., compute
\begin{align}
\vec\nabla_lz(\x)&=2(\vec x_l-\vec c)-\frac2N\sum_{j=1}^N(\vec x_j-\vec c)\,.
\end{align}
For the full $3N$-dimensional gradient we obtain
\begin{align}
\frac{[\nabla z(\x)]^2}4 &=z(\x)-\frac1{N}\sum_{j,l=1}^N(\vec x_j-\vec c)\cdot(\vec x_l-\vec c)\nonumber\\
&=z(\x)-\frac1{N}\left [\sum_{j=1}^N(\vec x_j-\vec c)\right ]^2\nonumber\\
&=z(\x)\,.
\end{align}
This yields the transport bound for $R_g^2$ in Eq.~(6) in the Letter. To further generalize, note that for time-dependent $\gamma(\tau)$, the factor $1/\gamma$ has to be kept inside the time integral. Moreover, note that if the beads have different $\gamma_j$ we may formulate the bound in terms of the minimum of these $\gamma_j$.

\subsection{Structure factor}
Consider the following $z_\tau$ and its derivative to approach $S_t(\vec q)$ and the corresponding $\mathcal D^z(t)$,
\begin{align}
z(\x) &\equiv \frac1N\sum_{j,k=1}^N\cos[\vec q\cdot(\vec x_j-\vec x_k)]\nonumber\\
S_t(\vec q)&=\E{z(\x_t)}\nonumber\\
\vec\nabla_j z(\x) &= -\frac{2\vec q}N\sum_{k=1}^N\sin[\vec q\cdot(\vec x_j-\vec x_k)]\nonumber\\
A_N&\equiv \frac{N^2}{4\vec q^{\,2}}[\nabla z(\x)]^2 = \sum_{j,k,l=1}^N\sin[\vec q\cdot(\vec x_j-\vec x_k)]\sin[\vec q\cdot(\vec x_j-\vec x_l)]\,.
\end{align}
In order to express the derivative in observable terms, consider the expression 
\begin{align}
B_N &\equiv \frac{N^2-Nz_{2\vec q}(\x)}4 = \frac14\sum_{k,l=1}^N\left(1-\cos[2\vec q\cdot(\vec x_k-\vec x_l)]\right)
=\frac12\sum_{j,k=1}^N\sin^2[\vec q\cdot(\vec x_k-\vec x_l)]\,.
\end{align}
Introduce the notation $s_{jk}=-s_{kj}=\sin[\vec q\cdot(\vec x_j-\vec x_k)]$ and note that from relabeling indices we get
\begin{align}
6(NB_N-A_N) &= 3N\sum_{k,l=1}^Ns_{kl}^2-6\sum_{j,k,l=1}^N s_{jl}s_{jk}
\nonumber\\
&=\sum_{j,k,l=1}^N\left(s_{jl}^2+s_{lk}^2+s_{kj}^2+2s_{jl}s_{lk}+2s_{lk}s_{kj}+2s_{kj}s_{jl}\right)\nonumber\\
&=\sum_{j,k,l=1}^N(s_{jl}+s_{lk}+s_{kj})^2\nonumber\\
&\ge 0\,,
\end{align}
with equality for $q\to 0$. This implies an upper bound of $t\mathcal D^z=4\vec q^{\,2}\int_0^t\rmd\tau\E{A_N(\tau)}/N^2\le 4\vec q^{\,2}\int_0^t\rmd\tau\E{B_N(\tau)}/N$ that implies the vector-version of the bound 
\begin{align}
T\DS\ge\frac{\gamma[S_t(\vec q)-S_0(\vec q)]^2}{\vec q^{\,2}\int_0^t[N-S_t(2\vec q)]\rmd\tau}\label{Sq bound vector} 
\end{align}
Note that this bound can be improved by using that we consider 3d space without assuming isotropy. Namely, by considering the above derivation for the observable 
\begin{align}
z_q(\x) &\equiv \frac13\left[z_{\vec q_1}(\x)+z_{\vec q_2}(\x)+z_{\vec q_3}(\x)\right]\nonumber\\
S_t(q) &\equiv \frac13\left[S_t(\vec q_1)+S_t(\vec q_2)+S_t(\vec q_3)\right]\,,
\end{align}
where we choose $\vec q_i=q\vec e_i$ where the $\vec e_i$ span an orthonormal basis [e.g., the vectors $(1,0,0)^T$, $(0,1,0)^T$, $(0,0,1)^T$], i.e., we average the structure factor over the three spatial dimensions. This allows to derive the extra factor of $3$ compared to Eq.~\eqref{Sq bound vector} to obtain the bound in Eq.~(6) in the Letter,
\begin{align}
T\DS\ge\frac{\gamma[S_t(\vec q)-S_0(\vec q)]^2}{\vec q^{\,2}\int_0^t[N-S_t(2\vec q)]\rmd\tau}\,,
\label{Sq bound scalar} 
\end{align}
since the sum of the $\vec q_i$ values in $(\nabla z)^2$ gives rise to an extra factor of 3 originating from $[(\vec q_1+\vec q_2+\vec q_3)/3]^2=(1,1,1)\cdot(1,1,1)/9=1/3$.

As for $R_g^2$, to further generalize, note that for time-dependent $\gamma(\tau)$, the factor $1/\gamma$ has to be kept inside the time integral. Moreover note that if the beads have different $\gamma_j$ we may formulate the bound in terms of the minimum of these $\gamma_j$.

Note that we can obtain the $R_g$-bound in Eq.~(6) in the Letter from Eq.~\eqref{Sq bound scalar} using Guinier's law for $q\to 0$, which means that we can obtain the value of the $R_g$-bound from the $S_t(q)$-bound (see $q\to 0$ in Fig.~2c,e in the Letter).

\section{Details on the example in Fig.~2 in the Letter}
\subsection{Computing the structure factor}
The time-resolved structure factor $S_t(q)$ and the corresponding quality factor of the bounds for the examples considered in Fig.~2 in the Letter are computed using normal mode analysis (the time integral in the denominator of the bound and integrals that averaged over rotated lattices are computed numerically).

The overdamped Rouse model considered in Figs.~2b,c in the Letter can be written as a $3N$ dimensional Langevin equation for the vector $\f r=(\vec r^{\,1},\dots,\vec r^{\,N})$ that reads (see \cite{SM_Dieball2022NJP,SM_Dieball2023PRR} for similar calculations)
\begin{align}
\rmd\f r_t=-\kappa\f k\f r_t\rmd t+\sqrt{2D}\rmd\f W_t,\label{Rouse first equation} 
\end{align}
with $D=k_{\rm B}T/\gamma$, trap stiffness $\kappa$, and the connectivity matrix $\f k$ is a $3N\times 3N$ matrix that reads ($\mathbbm{1}_3$ is the 3d unit matrix and all terms not shown are $0$)
\begin{align}
\f k=\begin{bmatrix}\mathbbm 1_3&-\mathbbm 1_3&&&&&\\-\mathbbm 1_3&2\mathbbm 1_3&-\mathbbm 1_3&&&0&\\&-\mathbbm 1_3&2\mathbbm 1_3&-\mathbbm 1_3&&&\\&&&\dots&\dots&&\\&&&&\dots&\dots&\\&0&&&-\mathbbm 1_3&2\mathbbm 1_3&-\mathbbm 1_3\\&&&&&-\mathbbm 1_3&\mathbbm 1_3\end{bmatrix}\,.
\end{align}
To model the nano-crystal, we consider the same harmonic model (i.e.,
linear drift) but for the deviations $\delta\vec r^{\,j}$ from the
rest positions, adapt the connectivity as shown in the schematic in Fig.~2a in the Letter (see also \cite{SM_Dieball2022NJP} on how to adapt connectivity), and also add a confinement by adding $\kappa_{\rm confine}\mathbbm 1_{3N}$ to $\kappa\f k$ in the drift term that keeps the beads closer to the prescribed mean positions, and avoids center-of-mass diffusion of $\delta\f R$.

The models are solved via normal mode analysis, i.e., by working in the basis that diagonalizes $\f A=\kappa\f k$ (or for the nano-crystal $\f A=\kappa\f k+\kappa_{\rm confine}\mathbbm 1_{3N}$).

The parameters chosen for Fig.~2 in the Letter in units of $k_{\rm B}=1$ are in (b,c) stiffness $\kappa=1$, $D=1$, quench from $T_0=2$ to $T=1$, and in (d,e) stiffness $\kappa=10$, confinement $\kappa_{\rm confine}=10$, rest positions spaced by $1$ on cubic lattice and quenched to rectangular lattice with spacings $1, 1.2, 0.7$ in $x,y,z$-direction.

Now change the basis to work in normal modes, i.e., define $\f Q$ and $\f X=\f Q^T\f R$ that fulfill $\f Q^T\f A\f Q={\rm diag}(a_l)$, such that we have $\rmd\vec X_t^l=-a_l\vec X_t^l\rmd t+\sqrt{2D}\rmd\vec W_t^{l}$ where the $W_t^{l,i}$ are mutually independent Wiener processes (see also Appendix F of \cite{SM_Dieball2022NJP}). This orthogonal change of basis exists since $\f A$ is symmetric.

For Gaussian displacements we have (see also \cite{SM_Dieball2022NJP})
\begin{align}
S_t(\vec q)
&\equiv\frac1N\sum_{j,k=1}^N\Elr{\rme^{-\rmi\vec q\cdot(\vec r_t^{\,j}-\vec r_t^{\,k})}}
=\frac1N\sum_{j,k=1}^N\exp\left[-\rmi \vec q\cdot\Elr{\vec r_t^{\,j}-\vec r_t^{\,k}}-\frac{\vec q^{\,2}}6\var\left(\vec r_t^{\,j}-\vec r_t^{\,k}\right)\right]\,.
\end{align}
Consider positions of beads $\vec r^{\,j}=\overline{\vec
  r^{\,j}}+\delta \vec r^{\,j}$ with a fixed (deterministic) rest position $\overline{\vec r^{\,j}}$ and random fluctuations $\delta \vec r^{\,j}$. This includes the Rouse model since if we set $\overline{\vec r^{\,j}}=0$ and $\kappa_{\rm confine}=0$ (then we will have center-of-mass diffusion but this can be neglected in the comoving frame). Using properties of Langevin dynamics with linear drift in the $\f Q$-rotated basis (see Appendix F of \cite{SM_Dieball2022NJP}), we obtain for the Rouse chain where $\Elr{\vec r_t^{\,j}-\vec r_t^{\,k}}=0$ and we quench the temperature (entering the variance) that
\begin{align}
\vec r_t^{\,j}&=\sum_{\alpha=1}^N Q_{j\alpha}\vec X_t^\alpha
\nonumber\\
\rmd\vec X_t^\alpha&=-a_\alpha\vec X_t^\alpha\rmd t+\sqrt{2D}\rmd\vec W_t^{\alpha}
\nonumber\\
V_\alpha(t)\equiv\E{(\vec X^\alpha_t)^2}&=\frac{3D}{a_\alpha}\left[1+\left(\frac{T_0}T-1\right)\rme^{-2a_\alpha t}\right]\quad\textnormal{for }\alpha\ne1
\nonumber\\
\E{\vec X^\alpha_t\cdot\vec X^\varphi_t}&=0\quad\textnormal{for }\alpha\ne\varphi
\nonumber\\
\Elr{(\vec r_t^{\,j})^2}&=\sum_{\alpha=1}^N V_\alpha(t)Q_{j\alpha}^2
\nonumber\\
\Elr{(\vec r_t^{\,j}-\vec r_t^{\,k})^2}&=\sum_{\alpha=2}^N V_\alpha(t)\left[Q_{j\alpha}^2+Q_{k\alpha}^2-2Q_{j\alpha}Q_{k\alpha}\right]
\nonumber\\
S_t(\vec q)&\equiv\frac1N\sum_{j,k=1}^N\Elr{\rme^{-\rmi\vec q\cdot(\vec r_t^{\,j}-\vec r_t^{\,k})}}=\frac1N\sum_{j,k=1}^N\rme^{-\vec q^{\,2}\Elr{(\vec r_t^{\,j}-\vec r_t^{\,k})^2}/6}
\,.\label{NMA temperature quench} 
\end{align}
On the other hand, for the nano-crystal where we do not quench
temperature (and thus also not the variances), but instead the rest
positions $\{\overline{\vec{r}^j}\}_{j\in[1,N]}$ (which we write as a
vector in $3N$-dimensional space $\overline{\f R})$, we have
\begin{align}
\delta\vec r_t^{\,j}&=\sum_{\alpha=1}^N Q_{j\alpha}\vec X_t^\alpha
\nonumber\\
\vec\mu_\alpha(t)\equiv\E{\vec X^\alpha_t} &= \rme^{-a_\alpha t}\vec\mu_\alpha(0)\quad\textnormal{for all }\alpha
\nonumber\\
\Elr{\vec r_t^{\,j}-\vec r_t^{\,k}} &= \overline{\vec r^{\,j}}-\overline{\vec r^{\,k}} + \Elr{\delta\vec r_t^{\,j}-\delta\vec r_t^{\,k}}
\nonumber\\
-\rmi\vec q\cdot\Elr{\delta\vec r_t^{\,j}-\delta\vec r_t^{\,k}} &= \sum_{\alpha=1}^N (Q_{j\alpha}-Q_{k\alpha})\rme^{-a_\alpha t}\vec\mu_\alpha(0)
\nonumber\\
\boldsymbol\mu(0) &= Q^T(\overline{\f R}_{\rm before\ quench}-\overline{\f R})
\nonumber\\
\var\left(\delta\vec r_t^{\,j}-\delta\vec r_t^{\,k}\right) &=
\var\left(\sum_{\alpha=1}^N (Q_{j\alpha}-Q_{k\alpha})\vec X_t^{\,\alpha}\right) 
\nonumber\\ &= \sum_{\alpha,\varphi=1}^N(Q_{j\alpha}-Q_{k\alpha})(Q_{j\varphi}-Q_{k\varphi})\cov_{\rm scalar}\left(\vec X^\alpha,\vec X^\varphi\right)
\nonumber\\
V_\alpha(t)\equiv\var\left(\vec X^\alpha_t\right) &= \frac{3D}{a_\alpha}\left[1+\left(\frac{V_\alpha(0)}{3D/a_\alpha}-1\right)\rme^{-2a_\alpha t}\right]\quad\textnormal{for all }\alpha
\nonumber\\
V_\alpha(0) &= 3D/a_\alpha\textnormal{ for structure quench, such that here }V_\alpha(t)=V_\alpha(0)
\nonumber\\
\cov_{\rm scalar}\left(X^\alpha_t,\vec X^\varphi_t\right) &= \delta_{\alpha\varphi}V_\alpha(t) 
\nonumber\\
\Rightarrow\quad \var\left(\delta\vec r_t^{\,j}-\delta\vec r_t^{\,k}\right) &= \sum_{\alpha=1}^N V_\alpha(t)\left[Q_{j\alpha}^2+Q_{k\alpha}^2-2Q_{j\alpha}Q_{k\alpha}\right]
\nonumber\\
S_t(\vec q) &= \frac1N\sum_{j,k=1}^N\exp\left[-\rmi\vec q\cdot\left(\overline{\vec r^{\,j}}-\overline{\vec r^{\,k}}\right)\right]\exp\left[-\rmi \vec q\cdot\Elr{\delta\vec r_t^{\,j}-\delta\vec r_t^{\,k}}-\frac{\vec q^{\,2}}6\var\left(\delta\vec r_t^{\,j}-\delta\vec r_t^{\,k}\right)\right]\nonumber\\
&= \frac1N\sum_{j,k=1}^N\cos\left[\vec q\cdot\left(\overline{\vec r^{\,j}}-\overline{\vec r^{\,k}}+\Elr{\delta\vec r_t^{\,j}-\delta\vec r_t^{\,k}}\right)\right]\exp\left[-\frac{\vec q^{\,2}}6\var\left(\delta\vec r_t^{\,j}-\delta\vec r_t^{\,k}\right)\right]\,.\label{NMA crystal quench} 
\end{align}
This is numerically implemented for the described quench in $\overline{\vec r^{\,j}}$, and afterwards, averaged over all directions since the crystals are randomly oriented in the solution (which is equivalent to an average over all directions of $\vec q$)
\begin{align}
S_t(q)=\frac1{4\pi}\int_{\abs{\vec q}=q}S_t(\vec q)\,.
\end{align}
From these results for $S_t(q)$ we compute the bounds in Eq.~(6) in the Letter. We could directly compute $R_g^2$ from $\E{(\vec r_t^{\,j}-\vec r_t^{\,k})^2}$ from the normal mode analysis, but instead we simply determine the value of the $R_g^2$ bound form the $q\to 0$ limit of the $S_t(q)$ bound.

\subsection{Computation of entropy production for the temperature quench}
To evaluate the quality factor $Q$, we need to divide by the bound by
the actual entropy production. As a shortcut to computing the entropy
production we may use that the transient entropy production of systems approaching an equilibrium state is given by the difference in excess free energy (since in this case, the total entropy production equals the non-adiabatic entropy production, see \cite{SM_Dieball2023PRR}), which in turn is computed from a Kullback-Leibler divergence \cite{SM_Lapolla2020PRL}. 

Note that we do not consider the entropy production of the center-of-mass diffusion ($\alpha=1$ mode) since it does not enter $S_t(\vec q)$, and since we may assume that this diffusion is already equilibrated in the sample (in which case it does not produce any entropy). Thus, we have for the excess free energy \cite[Eq.~(6)]{SM_Lapolla2020PRL}
\begin{align}
{\rm EFE}(t) &= \frac{3k_{\rm B}}2\sum_{\alpha=2}^N\left\{\left(\frac{T_0}T-1\right)\rme^{-2a_\alpha t}-\ln\left[1+\left(\frac{T_0}T-1\right)\rme^{-2a_\alpha t}\right]\right\}
\nonumber\\
&= \frac{3k_{\rm B}}2\sum_{\alpha=2}^N\left\{\left(\frac{T_0}T-1\right)\rme^{-2a_\alpha t}-T\ln\left[1+\left(\frac{T_0}T-1\right)\rme^{-2a_\alpha t}\right]\right\}\nonumber\\
\DS[0,t] &=\DS[0,\infty]-\DS[t,\infty]= {\rm EFE}(0)-{\rm EFE}(t)\,.\label{EFE Rouse} 
\end{align}

\subsection{Computation of entropy production for the structure quench}
Again, the entropy production is computed from the Kullback-Leibler
divergences, but this time for Gaussians with the same variance but
different mean values, where we have
\begin{align}
2D_{KL}[P(t)||P(\infty)] = (\boldsymbol\mu_t-\boldsymbol\mu_\infty)^T\bSig_\infty^{-1}(\boldsymbol\mu_t-\boldsymbol\mu_\infty)
\,.\label{DKL Gauss constant covariance} 
\end{align}
For the considered quench we have, see Eq.~\eqref{NMA crystal quench}, (note that we need to include $\alpha=1$ here since $\kappa_{\rm confine}\ne0$ implies that there is no center-of-mass diffusion)
\begin{align}
\boldsymbol\mu_\infty &= \f 0
\nonumber\\
\boldsymbol\mu_0 &= Q^T(\overline{\f r}_{\rm before\ quench}-\overline{\f r})
\nonumber\\
\boldsymbol\mu_t &= \exp(-{\rm diag}(\f a)t)\boldsymbol\mu_0
\nonumber\\
\bSig_\infty^{-1} &= {\rm diag}(\f a)/D
\nonumber\\
{\rm EFE}(t)/T=D_{KL}[P(t)||P(\infty)] &= \frac1{2D}\sum_{\alpha=1}^N a_\alpha\exp(-2a_\alpha t)\vec\mu_\alpha(0)^2
\nonumber\\
\Delta S_{\rm tot} &= {\rm EFE}(0)/T-{\rm EFE}(t)/T=\frac1{2D}\sum_{\alpha=1}^N a_\alpha[1-\exp(-2a_\alpha t)]\vec\mu_\alpha(0)^2
\,.
\end{align}

\section{Details on the handling of $D^z$}
If we measure the full dynamics $x_\tau$ and know, or even choose,  $z_\tau$, then we can directly evaluate $\mathcal D^z$.
Moreover, if we know $z_\tau$ but do \emph{not} measure $x_\tau$, we can often express $\mathcal D^z$ in terms of $\E{z_\tau}$, or bound it by this and constants (as for $R_g^2$ and $S_t(q)$ above).  

However, in the challenging case that we only measure $z_\tau$ but do
\emph{not} know what function of the underlying dynamics $x_\tau$ it
is (this would, e.g., be the case if we measure the dynamics of
a macromolecule along some reaction
coordinate), then we can often obtain $\mathcal D^z$ from the short-time fluctuations of $z_\tau$. 

\subsection{Overdamped dynamics}
For overdamped dynamics, we obtain $\mathcal D^z(t)$ directly from the from short-time fluctuations of $z_\tau$ (see also Ref.~\cite{SM_Dechant2023PRL}),
\begin{align}
\rmd z_\tau&=\nabla_\x z_\tau\cdot\rmd\x_\tau+\dot z_\tau\rmd t=k_{\rm B}T\nabla_\x z_\tau\cdot\bgam^{-1}\rmd\f W_\tau+O(\rmd\tau)\nonumber\\
k_{\rm B}T\Elr{\nabla_\x z_\tau\cdot\bgam^{-1}\nabla_\x z_\tau} &= \frac{\var[\rmd z_\tau]}{2\rmd\tau}\nonumber\\
k_{\rm B}Tt\mathcal D^z(t) &= \int_0^t\rmd\tau\frac{\var[\rmd z_\tau]}{2\rmd\tau}\,.
\end{align}

\subsection{Underdamped dynamics}
For simplicity, assume $\gamma$ and $m$ to be scalars and $z_\tau$ to not explicitly depend on time. Then, we can write
\begin{align}
\frac{\rmd}{\rmd \tau} z(\x_\tau) &=\nabla_\x z(\x_\tau)\cdot\f v_\tau 
\nonumber\\
\rmd\left[\frac{\rmd}{\rmd \tau} z(\x_\tau)\right] &=\nabla_\x z(\x_\tau)\cdot\rmd \f v_\tau+\rmd[\nabla_\x z(\x_\tau)]\cdot\f v_\tau
\nonumber\\
&=\nabla_\x z(\x_\tau)\cdot\frac{\sqrt{2k_{\rm B}T\gamma}}{m}\rmd \f W_\tau+O(\rmd t)
\nonumber\\
\frac{\var\left(\rmd\left[\frac{\rmd}{\rmd \tau}
    z(\x_\tau)\right]\right)}{2k_{\rm B}T\rmd t}&=\left(\frac{\gamma}{m}\right)^2\Elr{\nabla_\x z(\x_\tau)\cdot\gamma^{-1}\nabla_\x z(\x_\tau)}\,.
\end{align}
Hence, if we know $\gamma/m$, or can determine it from another experiment, we can again obtain $t\mathcal D^z(t)$ from observations of $z_\tau$.

If we can observe $z_\tau$ while the system settles into a steady state, we can even infer $\gamma/m$ from $z_\tau$ itself. We here distinguish two cases. First, if the dynamics approach equilibrium, then we can obtain $\gamma/m$ from comparing the equilibrium fluctuations $\var_{\rm eq}(\rmd z)/\rmd t^2\sim \Elr{\nabla_\x z(\x_\tau)\cdot\gamma^{-1}\nabla_\x z(\x_\tau)}\gamma/m$ (see Supplemental Material of Ref.~\cite{SM_Dechant2023PRL}) to the previously mentioned equilibrium result for $\var\left(\rmd\left[\frac{\rmd}{\rmd \tau} z(\x_\tau)\right]\right)\sim \Elr{\nabla_\x z(\x_\tau)\cdot\gamma^{-1}\nabla_\x z(\x_\tau)}(\gamma/m)^2$.

Second, if approach a non-equilibrium steady state, we can still bound the thermalization time $\tau_{\rm thermalization}=m/\gamma$ since it cannot be faster than the relaxation time scale $\tau^{-1}_{\rm relax}=-\lim_{t\to\infty}t^{-1}\ln C_{zz}(t)\ge \tau_{\rm thermalization}$, where $C_{zz}(t)=\E{z_t z_0}-\E{z_t}\E{z_0}$

\section{FRET bound}
For the FRET efficiency $E_t$ measured between donor and acceptor
chromophores at positions
$\vec x_1$ and $\vec x_2$ with F\"orster radius $R_0$ we can compute
\begin{align}
E_t&=\Elr{\left[1+\left(\frac{\vec x_1-\vec x_2}{R_0}\right)^6\right]^{-1}}=\E{z(\x)}\nonumber\\
z(\x)&=\left[1+\left(\frac{\vec x_1-\vec x_2}{R_0}\right)^6\right]^{-1}\le 1\nonumber\\
\vec\nabla_1 z(\x)&=-z^2(\x)6\left(\frac{\vec x_1-\vec x_2}{R_0}\right)^5\frac{\vec\nabla_1\cdot\vec x_1}{R_0}=-\frac{2z^2(\x)}{R_0}\left(\frac{\vec x_1-\vec x_2}{R_0}\right)^5\\
[\nabla z(\x)]^2&=[\vec\nabla_1 z(\x)]^2+[\vec\nabla_2 z(\x)]^2=\frac{8z^4(\x)}{R_0^2}\left(\frac{\vec x_1-\vec x_2}{R_0}\right)^{10}=\frac{8z^4(\x)}{R_0^2}\left(\frac1{z(\x)}-1\right)^{5/3}=\frac{8z^{7/3}(\x)}{R_0^2}[1-z(\x)]^{5/3}\le\frac8{R_0^2}\,.\nonumber
\end{align}
Using this simple approximation, we obtain from the transport bound \eqref{transport bound SM}
\begin{align}
T\DS\ge R_0^2\gamma[E_t-E_0]^2/8t\,. \label{FRET bound SM} 
\end{align}
This bound can be easily improved, e.g., by bounding $z$ and $1-z$ above by the maximally measured values instead of by $1$.

\end{document}